\documentclass[aps,prc,showpacs,nofootinbib]{revtex4}

\usepackage{epsfig}
\usepackage{graphicx}

\begin{document}
\title{Polarization phenomena in open charm photoproduction processes}
\author{Egle Tomasi-Gustafsson}
\email{etomasi@cea.fr}
\affiliation{\it DAPNIA/SPhN, CEA/Saclay, 91191 Gif-sur-Yvette Cedex,
France}
\author{Michail P. Rekalo} 
\altaffiliation{Permanent address:
\it NSC Kharkov Institute of Physics and Technology, 61108 Kharkov, Ukraine}
\affiliation{Middle East Technical University,
Physics Department, Ankara 06531, Turkey}

\date{\today}

\begin{abstract}
We analyze polarization effects in associative photoproduction of pseudoscalar ($\overline{D}$) charmed mesons in exclusive processes $\gamma+ N\to Y_c +\overline{D}$, $Y_c=\Lambda_c^+$, $\Sigma_c$. Circularly polarized photons induce nonzero polarization of the $Y_c$-hyperon with $x$- and $z$-components (in the reaction plane) and 
non vanishing asymmetries ${\cal A}_x$ and ${\cal A}_z$ for polarized nucleon target. These polarization observables can be predicted in model-independent way for exclusive $\overline{D}$-production processes in collinear kinematics. The T-even $Y_c$-polarization and asymmetries for non-collinear kinematics can be calculated in framework of an effective Lagrangian approach. The depolarization coefficients $D_{ab}$, characterizing the dependence of the $Y_c$-polarization on the nucleon polarization are also calculated. 
\end{abstract}
\maketitle
\section{Introduction}

The experimental study of photoproduction of charmed particles on nucleons through exclusive and inclusive reactions:  $\gamma+N\to Y_c+\overline{D_c} $ $(\overline{D_c^*}) $, $Y_c=\Lambda_c$($\Sigma_c$), 
$\gamma+N\to N +D_c+\overline{D_c}$, 
$\gamma+N\to \Lambda_c(\overline{\Lambda_c})+X$ started about twenty years ago, \cite{As80,Ad87}, in the energy range  $E_{\gamma}= 20\div 70 $ GeV.  Since then, several experiments were devoted to this subject, \cite{Al92,Anjos89,Au83}, but, up to now, the smallest energy where data are available is $E_{\gamma}$=20 GeV,  \cite{Ab86}, which is still relatively far from threshold (for example, $E_{thr}=8.7$ GeV for $\gamma+p\to \Lambda_c^+ +\overline{D^0}$).

Charm particle photoproduction is usually interpreted according to the photon-gluon fusion mechanism (PGF), $\gamma+G\to c+\overline{c}$ \cite{PGF}, which can be considered as the simplest QCD hard process (called leading order (LO) approximation).

Considering the corresponding fragmentation functions for $c\to D_c+X$, $c\to Y_c+X$, $c\to\overline{D_c}+X$..., the observables for inclusive processes of charmed mesons and hyperons photoproduction can be calculated. The  existing experimental data on the total cross section for open charm photoproduction, $\gamma+N\to charm+X$, the relative production of $D^0$ and $D^+$-mesons as well as $\Lambda_c^+$--hyperon are usually interpreted within this approach.

In principle, other mechanisms (of non-perturbative nature) should also be taken into account, such as, for example, the diffractive production of $D\overline{D}$ or $\Lambda_c D$ pairs, through Pomeron exchange. Different hadronic exchanges can also contribute to the simplest exclusive processes, $\gamma+N\to Y_c+ \overline{D_c} $ $(\overline{D_c^*}) $, in the near threshold region \cite{Re1,Re2}. More experimental and theoretical studies are needed in order to clarify the physics of charm particle photoproduction.

Up to now, all charm photoproduction experiments have been done with unpolarized particles\footnote{We should mention here an earlier attempt \protect\cite{Ab86} to study open charm photoproduction with linearly polarized photons.}. Polarization phenomena will be essential for the understanding of the reaction mechanism, in particular the importance of the main subprocess, 
$\gamma+G\to c+\overline{c}$. For example, T-even polarization observables as the $\Sigma _B$-asymmetry \cite{Du80}, induced by a linearly polarized photon beam, the asymmetries in collisions of circularly polarized photons on polarized gluons \cite{Wa82} and the polarization of the $c$-quark have relatively large absolute values, in framework of this model and can be actually measured. The running COMPASS experiment \cite{COMPASS}, with longitudinally polarized muons and polarized targets ($p$ or $LiD$) will access these polarization effects.

High energy photon beams with large degree of circular polarization are actually available for physical experiments. Complementary to the linear polarized photon beams, they allow to address different interesting problems of hadron electrodynamics. Circularly polarized photon beams can be obtained in different ways, for example by backward scattering of a laser beam by high energy electron beam with longitudinal polarization. In JLab circularly polarized bremsstrahlung photons were generated by polarized electrons \cite{pol1,pol2}. The  proton polarization in the process of deuteron photodisintegration, $\vec \gamma+d\to \vec p+n$ \cite{Wi01} was investigated to test QCD applicability \cite{Br81}, with respect to hadron helicity conservation in high-energy photon-deuteron interactions. Coherent scattering of a longitudinally polarized electron beam by a diamond crystal results in a very particular beam of circularly polarized photons \cite{Gh00}. In the COMPASS experiment \cite{COMPASS}, the high energy longitudinally polarized muon beam generates circularly polarized photons (real and virtual) in a wide energy interval (50-130 GeV). 

The interest of a  circularly polarized photon beam is related with the very actual question about how the proton spin is shared among its constituents  \cite{Ad94a,Ad95,Ab94a,Ab94b}. The determination of the gluon contribution, $\Delta G$, to the nucleon spin is the object of different experiments with polarized beams and targets \cite{Ad94b,Bu92}. 

In particular, the production of charmed particles in collisions of longitudinally polarized muons with polarized proton target will be investigated by the COMPASS collaboration. In the framework of the photon-gluon fusion model \cite{PGF}, $\gamma^*+G\to c+\overline{c}$, ($\gamma^*$ is a virtual photon), the corresponding asymmetry (for polarized $\mu p$-collisions), is related to the polarized gluon content in the polarized protons \cite{Br97,Gl88,Gs96}. Due to the future impact of such result, it seems necessary to understand all the other possible mechanisms which can contribute to inclusive charm photoproduction, such as $\gamma+p\to \overline{D^0}+X$, for example. One possible and $a~priori$ important background, which can be investigated in detail, is the process of exclusive associative charm photoproduction, with pseudoscalar and vector charmed mesons in the final state,  $\gamma+p\to Y_c+\overline{D}(\overline{D^*})$. The mechanism of photon-gluon fusion, which successfully describes the inclusive spectra of $D$ and $D^*$ mesons at high photon energies, can not be easily applied to exclusive processes, at any energy. Threshold considerations of such processes in terms of standard perturbative QCD have been applied only to the energy dependence of the cross section of $\gamma+p\to p+J/\psi$ \cite{Br00}, as, in such approach, polarization phenomena can not be calculated without additional assumptions.

In this respect, the formalism  of the effective Lagrangian approach (ELA) seems very convenient for the calculation of exclusive associative photoproduction of charmed particles, such as $\gamma+N\to Y_c^+\overline{D}(\overline{D^*})$ \cite{Re1,Re2,Re02} at least in the near threshold region. Such approach is also widely used for the analysis of various processes involving charmed particles \cite{Li00}, as, for example, $J/\psi$-suppression in high energy heavy ion collisions, in connection with quark-gluon plasma transition 
\cite{Ma86}. We analyzed earlier different exclusive processes of associative charm particles photoproduction, $\gamma+N\to Y_c^+ +\overline{D}(\overline{D^*})$ in the threshold region, and indicated that polarization phenomena can be naturally predicted for those reactions. In this paper we extend this model to higher energies by including the mechanism of $D^*$ exchange and apply the ELA approach to the collision of circularly polarized photons with an unpolarized and a polarized proton target. We predict the angular and energy dependences 
of the asymmetries in $\vec\gamma+\vec N\to Y_c^++\overline{D}$ processes and of the polarizations of the $Y_c^+$-hyperons, produced in $\vec\gamma + N\to Y_c^++\overline{D}$ with circularly polarized photons (on polarized target) and in $\gamma +\vec N\to Y_c^++\overline{D}$ (with unpolarized photons on polarized target).

The exclusive reactions, $\gamma+N\to Y_c+\overline{D}$ $ (\overline{D^*})$, which are object of this paper, are important not only as possible background for experiments aiming to the measurement of the gluon contribution to the nucleon spin, but these processes have an intrinsic physical interest, in the understanding of charm particle electrodynamics. Let us mention some important aspects:
\begin{itemize}
\item possibility to measure the electromagnetic properties of charmed particles, such as the magnetic moments of $Y_c$-hyperons or the vector $D^*$-meson;
\item test of $SU(4)$-symmetry for the strong $ND_cY_c$-coupling constants;
\item determination of the ${\cal P}$-parity of charmed particles;
\item explanation of the asymmetric ratio $D/ \overline{D}$,   $\overline{\Lambda_c}/\Lambda_c$ in open charm photoproduction reactions;
\item understanding of nonperturbative mechanisms for associative charm photoproduction.
\end{itemize}

The exclusive reactions $\gamma+N\to Y_c+\overline{D}$ $ (\overline{D^*})$ should largely contribute to the total cross section, in particular in the near threshold region, in analogy with the processes of vector meson production, $K$ and $K^*$ production in $\pi N$ or $NN$ collisions. For example, at $E_{\gamma}=20$ GeV, the associative $\Lambda_c\overline{D}$-production contribute 71\% to the total cross section of open charm photoproduction \cite{Ab86}.

The paper is organized as follows. Using the standard parametrization of the spin structure for the matrix element of $\gamma+N\to Y_c+\overline{D}$-processes, in Section II we calculate the one and two spin polarization observables in terms of four scalar amplitudes. Section III and IV contain a description of possible non-perturbative mechanisms for exclusive associative charm photoproduction and the relativistic parametrization of the corresponding matrix elements. In section V we give three set of parameters, corresponding to three possible models for $\gamma+p\to \Lambda_c^++\overline{D^0}$, which describe the energy dependence of the total cross section. Polarization phenomena for the three models are discussed in Section VI.

\section{Polarization observables for  $\gamma+ N\to Y_c+\overline{D}$}

We consider here different polarization observables for associative $Y_c\overline{D}$ photoproduction, starting from the simplest one-spin asymmetry $\Sigma_B$, induced by photons with linear polarization, to the depolarization coefficients $D_{ab}$, which describe the dependence of the polarization of the produced $Y_c$-hyperon on the nucleon target polarization.

\subsection{Spin structure of the matrix element and differential cross section}

We will use the standard parametrization \cite{Ch57} of the spin structure for the amplitude of pseudoscalar meson photoproduction on the nucleon:
\begin{equation}{\cal M}(\gamma N\to Y_c\overline{D})=\chi_2^{\dagger}\left [i\vec\sigma\cdot\vec e f_1 + 
\vec\sigma\cdot\hat{\vec q}\vec\sigma\cdot\hat{\vec k}\times\vec e f_2+ 
i\vec e \cdot\hat{\vec q}\vec\sigma\cdot\hat{\vec k}f_3
+i\vec\sigma\cdot\hat{\vec q} \vec e \cdot\hat{\vec q}f_4\right ]\chi_1,
\label{eq:mat}
\end{equation}
where $\hat{\vec k}$ and
$\hat{\vec q}$ are the unit vectors along the three-momentum of $\gamma$ and 
$\overline{D}$; $f_i$, $i$=1-4, are the scalar amplitudes, which are functions of two independent kinematical variables, the square of the total energy $s$ and $\cos\vartheta$, where $\vartheta$ is the $\overline{D}$--meson production angle in the reaction center of mass (CMS) (with respect to the direction of the incident photon),  $\chi_1$ and  $\chi_2$ are the two-component spinors of the initial nucleon and the produced $Y_c$-baryon. 

Note that the pseudoscalar nature of the $\overline{D}$--meson is not experimentally confirmed up to now, therefore we follow here the prescription of the quark model for the P-parities of $\overline{D}$ and $Y_c$-charm particles. 

The differential cross section when all particles are unpolarized, can be derived from (\ref{eq:mat}):
$$
\displaystyle\frac{d\sigma}{d\Omega}=\displaystyle\frac{q}{k}
\displaystyle\frac{(E_1+m)(E_2+M)}{64\pi^2s}{\cal N}_0$$
and
\begin{eqnarray}
&{\cal N}_0=&|f_1|^2+|f_2|^2-2\cos\vartheta {\cal R}e f_1f_2^*+ \nonumber\\
&&\sin^2\vartheta 
\left \{ \displaystyle\frac{1}{2}\left ( |f_3|^2+|f_4|^2 \right )+  
{\cal R}e \left [ f_2f_3^*+2\left (f_1+\cos\vartheta f_3 \right ) f_4^*\right ] \right  \}, 
\label{eq:asymd}
\end{eqnarray}
where $E_1(E_2)$ and $m(M)$ are the energy and the mass of $N(Y_c)$, 
respectively.
\subsection{Charm photoproduction with linearly polarized photons}

A linearly polarized photon beam, on an unpolarized proton target, may induce  a beam asymmetry $\Sigma_B$ defined as:
\begin{equation}
\Sigma_B=\displaystyle\frac{d\sigma_{\perp}/d\Omega-d\sigma_{\parallel}/d\Omega}
{d\sigma_{\perp}/d\Omega+d\sigma_{\parallel}/d\Omega},
\label{sigma}
\end{equation}
with 
\begin{equation} 
{\cal N}_0 \Sigma_B=-\displaystyle\frac{\sin^2}{2}\vartheta\left [ |f_3|^2+|f_4|^2+2{\cal R}e f_2f_3^*+2{\cal R}e(f_1+\cos\vartheta f_3)f_4^*\right ],
\label{sigma1}
\end{equation}
where $d\sigma_{\perp }/d\Omega $ and $d\sigma_{\parallel }/d\Omega$ are the differential cross sections for the absorption of photon with linear polarization, which is orthogonal or parallel to the reaction plane.

The measurement of this observable for the processes $\vec\gamma +N\to Y_c+\overline{D}$ is important as this reaction constitutes a possible background with respect to PGF, $\vec \gamma+G\to c+\overline{c}$. For all these processes, $\Sigma_B$ is the simplest polarization observable of T-even nature, which does not vanish in LO approximation. Moreover, for $\gamma+G\to c+\overline{c}$, it has a large sensitivity to the mass of the $c$-quark, as it is proportional to $m_c^2$ and it does not depend on the fragmentation functions, $c\to D(\Lambda_c)+X$. In this approach, $\Sigma_B(\vec \gamma N\to c\overline{c}X)$ is determined by the unpolarized gluon distribution $G(x)$, which is relatively well known, in comparison with the polarized gluon distribution, $\Delta G(x)$. So, in principle, $\Sigma_B(\vec \gamma N\to c\overline{c}X)$ can be predicted with better precision than the asymmetry $A_z(\vec \gamma \vec N\to c\overline{c}X)$ (in the collision of circularly polarized photons with a longitudinally polarized target) which is actually considered the most direct way to access $\Delta G(x)$.

High energy photon beams with linear polarization can be obtained at SLAC and at CERN. In the conditions of the COMPASS experiment, the scattering of muons, $p(\mu,\mu'\overline{D})X$, allows, in principle, to measure $\Sigma_B$ even for $Q^2\ne 0$. For this aim, it is necessary to study the $\phi$-dependence of the corresponding exclusive cross section, taking into account that the $\cos 2\phi$-contribution is proportional to $\Sigma_B$ (where $\phi$ is the azimuthal angle of $\overline{D}$-production, with respect to the muon scattering plane).

\subsection{Collisions of circularly polarized photons with a polarized proton target}

Circularly polarized photon beams allow to obtain additional dynamical information  in comparison with photons with linear polarization, in hadronic processes (photoproduction or photodisintegration). The difference between linear and circular photon polarizations is due to the P-odd nature of the photon helicity ($\lambda=\pm 1$), which is the natural characteristic of circularly polarized photons. Therefore, in any binary process, $\gamma+a\to b+c$ ($a$, $b$, and $c$ are hadrons and/or nuclei), with circularly polarized photons, the asymmetry, in case of unpolarized target, vanishes, due to the P-invariance of electromagnetic interactions of hadrons. In contrast, linearly polarized photons generally induce non-zero asymmetry, even in case of unpolarized target and unpolarized final particles. Circular polarization manifests itself only in two-spin (or more) polarization phenomena, as for example, the correlation polarization coefficient of photon beam and proton target, $\vec\gamma+\vec  p\to \Lambda_c^+ +\overline{D^0}$, or the polarization of the final   $\Lambda_c $-hyperon in $\vec\gamma+ p\to \vec \Lambda_c^+ +\overline{D^0}$. The analysis of these processes is the object of  the present paper. Note that in both cases,  the components of the target polarization and the $\Lambda_c^+$ polarization lie in the reaction plane, due to the P-invariance. Note also, that all these polarization observables are T-even, i.e. they may not vanish even if the amplitudes of the considered process are real.

Another possible two-spin correlation coefficient with circularly polarized photons for the process $\gamma+\vec  p\to \Lambda_c^+ +\overline{D^{*0}}$ is the vector polarization of the $D^*$-meson. In this connection, however, it is necessary to stress that the vector polarization of $D^*$ can not be measured through its  most probable decays: $D^*\to D+\pi$, $D^*\to D+\gamma$, $D^*\to D+e^+ + e^-$, which  are induced by P-invariant strong and electromagnetic interaction. The matrix elements for these decays are characterized by a single spin structure, which is insensitive to the vector polarization.

For completeness, let us mention two recent applications of high energy circularly polarized photon beams. One is the experimental verification of the Gerasimov-Drell-Hearn sum rule \cite{GDH}, which is determined by an integral of the difference of the total $\gamma N$-cross section, with two possible total spin projections, 3/2 and 1/2. Such quantity, $\sigma_{3/2}-\sigma_{1/2}$, can be measured in collisions of circularly polarized photons with a polarized nucleon target. Measurements are going on at MAMI, for $E_{\gamma}<800$ MeV \cite{gdhmami} and at JLab \cite{gdhjlab}.

In the general case, any reaction $\vec\gamma+\vec p\to
Y_c+\overline{D^0}(\overline{D^*})$ (with circularly polarized photons and polarized target) is described by two different asymmetries and the dependence of the differential cross section on the polarization states of the colliding particles can be parametrized as follows (taking into account the P-invariance of the electromagnetic interaction of charmed particles) \footnote{Note that we consider here only two-spin correlations, neglecting in particular T-odd analyzing power, induced by the $P_y$ component of the target polarization.}:
\begin{equation}
\displaystyle\frac{d\sigma}{d\Omega}(\vec\gamma\vec p)=\left ( \displaystyle\frac{d\sigma}{d\Omega}\right)_0
\left (1-\lambda T_x{\cal A}_x-\lambda T_z{\cal A}_z\right ),
\label{sig}
\end{equation}
where $\lambda =\pm 1$ is the photon helicity, $T_x$ and $T_z$ are the possible components of the proton polarization $\vec T$, ${\cal A}_x$ and ${\cal A}_z$ are the two independent asymmetries. Due to the T-even nature of these asymmetries, they are non vanishing in ELA consideration, where the photoproduction amplitudes $f_i$ are real. These asymmetries are nonzero also for hard QCD processes as PGF, $\gamma+G\to c+\overline{c}$, for collisions of polarized photons and gluons with definite helicities \cite{Wa82}.

\begin{figure}
\mbox{\epsfxsize=8.cm\leavevmode \epsffile{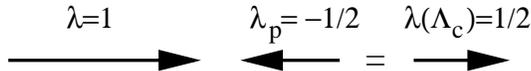}}
\vspace*{.2 truecm}
\caption{Conservation of helicity in  $\vec\gamma+\vec p\to \Lambda_c +D^0$, in collinear regime.}
\label{fig:heli}
\end{figure}

After summing over the $Y_c$-polarizations, one finds the following expressions for the asymmetries ${\cal A}_x$ and ${\cal A}_z$ in terms of 
the scalar amplitudes $f_i$:
\begin{equation}
{\cal A}_x {\cal N}_0=\sin\vartheta {\cal R}e \left [-f_1f_3^*+f_2f_4^*+\cos\vartheta 
(-f_1f_4^*+f_2f_3^*)\right ],
\label{asymx}
\end{equation}
\begin{equation}
{\cal A}_z{\cal N}_0={\cal R}e [|f_1|^2+|f_2|^2-2\cos\vartheta f_1f_2^*+ \sin^2\vartheta \left (f_1f_4^*+f_2f_3^*\right )].
\label{asymy}
\end{equation}
Note that  ${\cal A}_x$ vanishes at $\vartheta=0^0$ and $\vartheta=\pi$. Moreover ${\cal A}_z=1$, for $\vartheta=0^0$ and $\vartheta=\pi$, for any photon energy. This is a model independent result, which follows from the conservation of helicity in collinear kinematics. It is correct for any dynamics of the process, its physical meaning is that the collision of $\gamma$ and $p$ with parallel spins can not take place for collinear regime (Fig. \ref{fig:heli}). This result holds for any process of pseudoscalar and scalar meson photoproduction on a nucleon target (if the final baryon has spin 1/2). Such independence on the ${\cal P}$-parity of produced mesons is important.

For $\vartheta\ne 0$ and $\vartheta \ne \pi$ the results for ${\cal A}_x$ and ${\cal A}_z$ are model dependent. The models of photon-gluon fusion predict a value for the ${\cal A}_z$-asymmetry for the inclusive $\vec\gamma +\vec p\to \overline{D^{0}}+X$ process $\le$  30\%, at $E_\gamma\simeq$ 50 GeV, depending on the assumptions on the polarized gluon distribution, $\Delta G(x)$. Therefore, even a $10$ \% contribution of the exclusive process $\vec \gamma+\vec p\to \Lambda_c^++\overline{D^0}$ to the inclusive $\overline{D}$-cross section can induce a large correction to the photon-gluon fusion asymmetry, at forward angles. This should be taken into account in the extraction of $\Delta G$ from the asymmetry in the process $\vec \gamma+\vec p\to \overline{D^0}+X$. Evidently the processes here considered, $\gamma+p\to Y_c+\overline{D^0}$ do not contribute to the  $D^0$ production in the inclusive reaction $\vec \gamma+\vec N\to D^0+X$.

\subsection{ $\Lambda_c$-polarization in $\vec\gamma+N\to \vec\Lambda_c+\overline{D}$ }

In a similar way it is possible to consider the polarization properties of the
final $\Lambda_c^+$ hyperon induced by the initial circular polarization of the
photon. Again, due to the P-invariance of the electromagnetic hadron
interaction,  only the $P_x$ and $P_z$-components do not vanish. In terms of the scalar amplitudes $f_i$,  defined above, the components of the $\Lambda_c^+$ polarization can be written as:
$$P_x {\cal N}_0=\lambda \sin\vartheta {\cal R}e \left [
-2f_1f_2^* - f_1f_3^*+
\cos\vartheta 
\left (2|f_2|^2+f_2f_3^*-f_1f_4^*\right) +(2\cos^2\vartheta-1) f_2f_4^*\right ],
$$
\begin{eqnarray}
&P_z {\cal N}_0= \lambda & {\cal R}e \left [|f_1|^2-(1-2\cos^2\vartheta )|f_2|^2-2\cos\vartheta 
 f_1f_2^*+ \right .\nonumber \\
&&\left .\sin^2\vartheta  (f_1f_4^*- f_2f_3^*
-2\cos\vartheta f_2f_4^* )\right ].
\label{pola}
\end{eqnarray}
Comparing Eqs. (\ref{asymx}), (\ref{asymy}) and  Eq. (\ref{pola}) one can see that the observables $A_x$ and $A_z$, on one side, and $P_x$ and $P_z$ on another side, are independent in the general case of non-collinear kinematics and contain different physical information. Note also that $P_z=1$, in case of collinear kinematics ( $\cos\vartheta =1$), independently on the model, taken to describe the scalar amplitudes $f_i$. This rigorous result follows from the conservation of the total helicity in $\gamma+ p\to \Lambda_c^+ (\Sigma_c^+)+\overline{D^0}$, which holds in collinear kinematics. It means that  only collisions with particles with antiparallel spins in the entrance channel are allowed (see Fig. \ref{fig:heli})
i.e. the final $\Lambda_c$ hyperon is polarized along the direction of the spin of the initial photon. This result holds for any $Y_c+\overline{D}$-final state independently on the ${\cal P}$-parity of the $N\Lambda_cD$-system.

At $\vartheta=0^0$ or $\vartheta=\pi$, the observable $P_x$ vanishes. This follows from the axial symmetry of the collinear kinematics, where only one physical direction can be defined (along the $z$-axis). In such kinematical conditions the $x$- and $y-$ axis are arbitrary, therefore $P_x={\cal A}_x=0$. It is a very general result, also independent on the relative ${\cal P}$-parity P$(N\Lambda_cD)$.

In non-collinear kinematics, the behavior of $P_x$ and $P_z$ can be predicted only in framework of a model. 

The measurement of the $\Lambda_c^+$-polarization can be done similarly to the strange $\Lambda^0$ hyperon, because the  $\Lambda_c^+$, being the lightest charm baryon, can decay only through the weak interaction. However the  $\Lambda_c^+$ decays through many channels with different branching ratios and analyzing powers. Let us mention the two particle decay $\Lambda_c^+\to \Lambda_c+\pi^+$, which has a large decay asymmetry, $A=0.98\pm 0.19$, but a relatively small branching ratio, $B(\Lambda \pi)=(9.0\pm2.8)10^{-3}$.
The semileptonic $\Lambda_c^+$ decay: $\Lambda_c^+\to \Lambda^0+e^+ +\nu_c$ is characterized by a larger branching ratio, $B(\Lambda e\nu)=(2.1\pm 0.6)$\%, with relatively large  decay asymmetry (in absolute value): $A=-0.82^{+0.11}_{-0.007}$
\cite{PDG}. Note that the possibility to measure the $\Lambda_c^+$-polarization has been experimentally confirmed in hadronic collisions \cite{Al86}.
It is not the case for the $\Sigma_c$-hyperon. Its main decay,  $\Sigma_c \to \Lambda_c +\pi$, is due to the strong interaction. So, whereas the $\Lambda_c^+$ is a self-analyzing particle, the $\Sigma_c$ is similar, in this respect, to any baryon resonance, with strong or electromagnetic decays.

\subsection{Depolarization coefficients}

Let us consider for completeness other two-spin polarization observables, the  coefficients of polarization transfer ${\cal D}_{ab}$, from a  polarized proton target to a $\Lambda_c$-hyperon, which can be defined as 
$${\cal N}_0 {\cal D}_{ab}=\displaystyle\frac{1}{2}Tr{\cal F}\vec\sigma\cdot {\vec a} {\cal F}^{\dagger}\vec\sigma\cdot {\vec b}.$$
The  index $a$, $a=m,n,k$ refers to the component of the proton polarization, whereas the index $b$, $b=m,n,k$  refers to the component of the $\Lambda_c$ polarization. The unit vectors $\hat{\vec m}$, $\hat{\vec n}$, and $\hat{\vec k}$ are defined as follows: $\hat{\vec k}=\vec k/|\vec k|$, 
$\hat{\vec n} =\vec k\times \vec q/|\vec k\times \vec q|$, $\hat{\vec m}=\hat{\vec n} \times\hat{\vec k}$.

Only five among these coefficients are nonzero (for unpolarized photons):
\begin{eqnarray}
{\cal N}_0{\cal D}_{mm}&=&\displaystyle\frac{\sin^2\vartheta}{2}
{\cal R}e~\left [ (2\sin^2\vartheta-1)|f_4|^2- |f_3|^2+2 
+f_1f_4^*-2f_2f_3^*-2\cos\vartheta(2 f_2f_4^*- f_3f_4^*)\right ]
\nonumber \\
{\cal N}_0{\cal D}_{nn}&=&-\displaystyle\frac{\sin^2\vartheta}{2}
{\cal R}e~\left [|f_3|^2+|f_4|^2+2(f_1f_4^*+f_2f_3^*+\cos\vartheta f_3f_4^*)\right ]
\nonumber \\
{\cal N}_0{\cal D}_{kk}&=&-{\cal R}e~\left \{ |f_1|^2+(1-2\cos^2\vartheta)|f_2|^2+2\cos\vartheta f_1f_2^*
+\displaystyle\frac{\sin^2\vartheta}{2}\left [ 
|f_3|^2+(2\cos^2\vartheta-1)|f_4|^2+\right .\right .\nonumber \\
&&
\left . \left . 2 (f_2f_3^*-f_1f_4^*+\cos\vartheta( 2f_2f_4^* +f_3f_4^*)
\right ] \right \}, \label{dij} \\
{\cal N}_0 {\cal D}_{mk}&=&\sin\vartheta
{\cal R}e~[\sin^2\vartheta(\cos\vartheta|f_4|^2+f_3f_4^*)+2f_1f_2^*+f_1f_3^*
+(1-2\cos^2\vartheta)~f_2f_4^*+\nonumber \\
&&
\cos\vartheta(-2|f_2|^2+f_1f_4^*-f_2f_3^* ],
\nonumber \\
{\cal N}_0{\cal D}_{km}&=&\sin\vartheta
{\cal R}e~[ \sin^2\vartheta(\cos\vartheta|f_4|^2+ f_3f_4^*)
+ \cos\vartheta(f_1f_4^*-f_2f_3^*)+f_1f_3^*+\nonumber \\
&&
(1-2\cos^2\vartheta)f_2f_4^*].
\nonumber 
\end{eqnarray}
Note that ${\cal D}_{km}\ne {\cal D}_{mk}$, because the process $\gamma+N\to Y_c+ \overline{D}$ is asymmetrical with respect to initial and final baryons, so these coefficients contain different information about the amplitudes $f_i$ and therefore about the reaction mechanisms.

Comparing Eqs. (\ref{sigma1}) and (\ref{dij}) one can see that $\Sigma_B$=$D_{nn}$. This relation is valid in the general case, for any kinematical condition and any choice of the reaction mechanism. In case of positive relative ${\cal P}$-parity of the $D$-meson with respect to the $N\Lambda_c$-system, ${\cal P}(N\Lambda_c D)$, we find, instead, $\Sigma_B=-D_{nn}$. Therefore we can write $\Sigma_B=-{\cal P}(N\Lambda_c D)D_{nn}$ and suggest a possible and model independent method to measure ${\cal P}(N\Lambda_c D)$, through the test of the relative sign of these two observables.
\section{Discussion of the reaction mechanisms}

\begin{figure}
\begin{center}
\includegraphics[width=7cm]{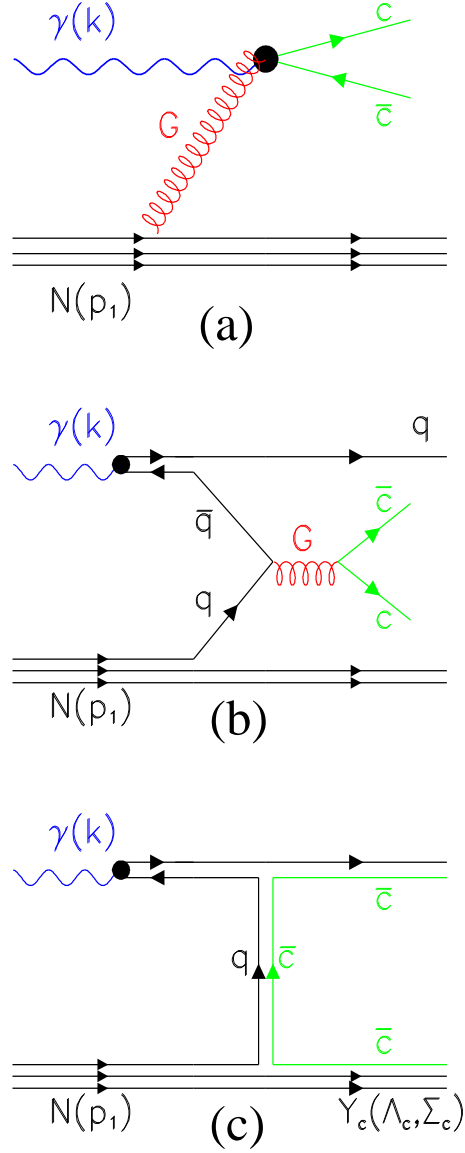}
\caption{\label{fig:fig1} Feynman diagrams for associative charm production in $\gamma N$-collisions: (a) the subprocess of photon-gluon fusion $\gamma+G\to c+\overline{c}$; (b) the subprocess of $q+\overline{q}$-annihilation, $q+\overline{q}\to c+\overline{c} $; (c) $t-$channel exchange by $\overline{D}$ and  $\overline{D}^*$ for $\gamma+N\to Y_c+\overline{D}(\overline{D^*})$.}
\end{center}
\end{figure}

The charm particle photo and electroproduction at high energy is usually interpreted in terms of photon-gluon fusion, $\gamma+G\to c+\overline{c}$ (Fig. \ref{fig:fig1}a). Near threshold, another possible mechanism, based on the subprocess $q+\overline{q}\to G\to c+\overline{c}$ (Fig. \ref{fig:fig1}b) should also be taken into account, as it was done for $\pi N$-collisions 
\cite{Sm97}. In case of exclusive reactions, $\gamma+N\to Y_c+\overline{D_c}$ $(\overline{D_c^*})$, $ Y_c=\Lambda_c,\Sigma_c$, the mechanism in Fig. \ref{fig:fig1}b  is equivalent to the exchange of a $\overline{c}q$-system, in $t$-channel (Fig. \ref{fig:fig1}c).  The importance of the annihilation mechanism for the explanation of forward charge asymmetry in $\gamma p$-collisions, has been investigated in \cite{Od87}.  So, one can find the mesonic equivalent of such exchange, i.e. the exchange by pseudoscalar $\overline{D_c}$ and (or) vector $\overline{D_c^*}$ mesons, in the $t$-channel of the considered reaction.

\begin{figure}
\begin{center}
\includegraphics[width=15cm]{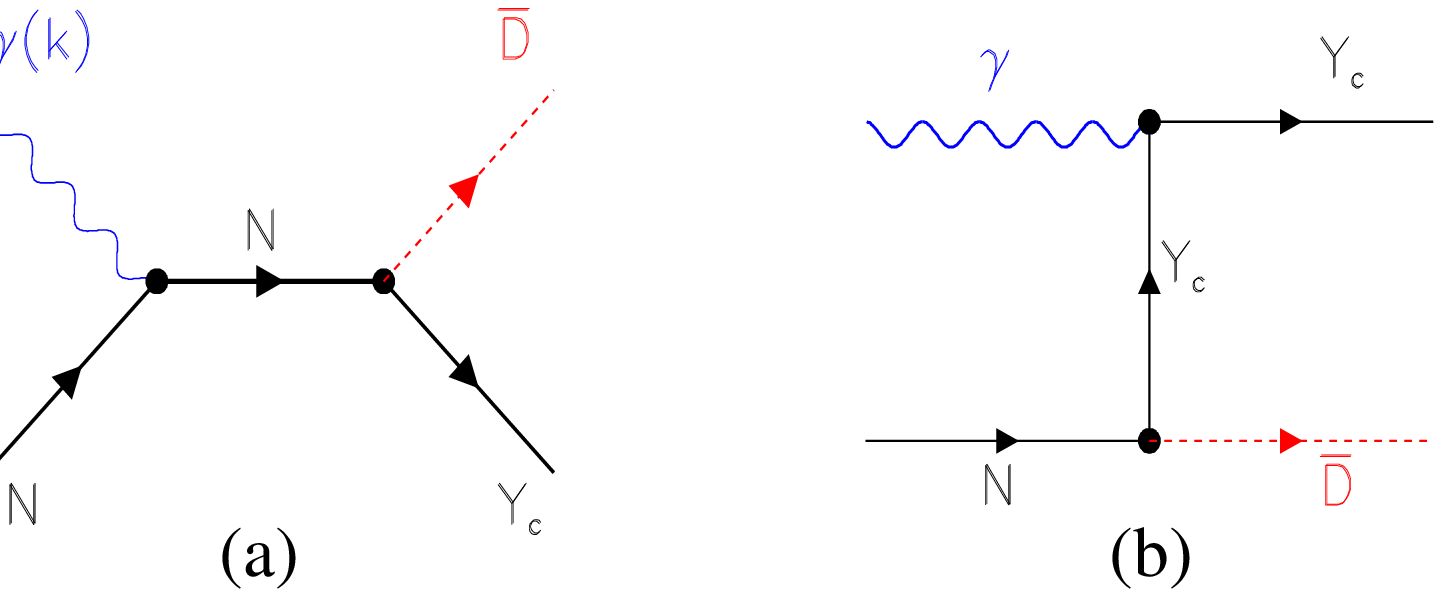}
\caption{\label{fig:fig2} Baryon exchanges in $s$-channel: (a), and $u$-channel: (b), for the process $\gamma+N\to Y_c+\overline{D}$.}
\end{center}
\end{figure}

To move further, one has to take into account the symmetry property of electromagnetic hadronic interaction, such as the conservation of electromagnetic current or the gauge invariance. This makes the principal difference between $\gamma N$ and $\pi N$ production of charmed particles. If in case of $D^*$ exchange in $\gamma+N\to Y_c+\overline{D_c}$ the corresponding matrix element is gauge invariant, for any kinematical conditions and any values of the coupling constants, due to the magnetic dipole transition in the vertex $D^*\to D+\gamma$, it is not the case for the pseudoscalar $D^-$ exchange in the reaction $\gamma+p\to \Sigma_c^{++}+D^-$, for example. The $D^-$ exchange alone can not satisfy the gauge invariance, therefore other mechanisms have to be added, such as baryonic exchanges in $s$- and $u$-channels (Fig. \ref{fig:fig2}).

\begin{figure}[h]
\begin{center}
\includegraphics[width=5cm]{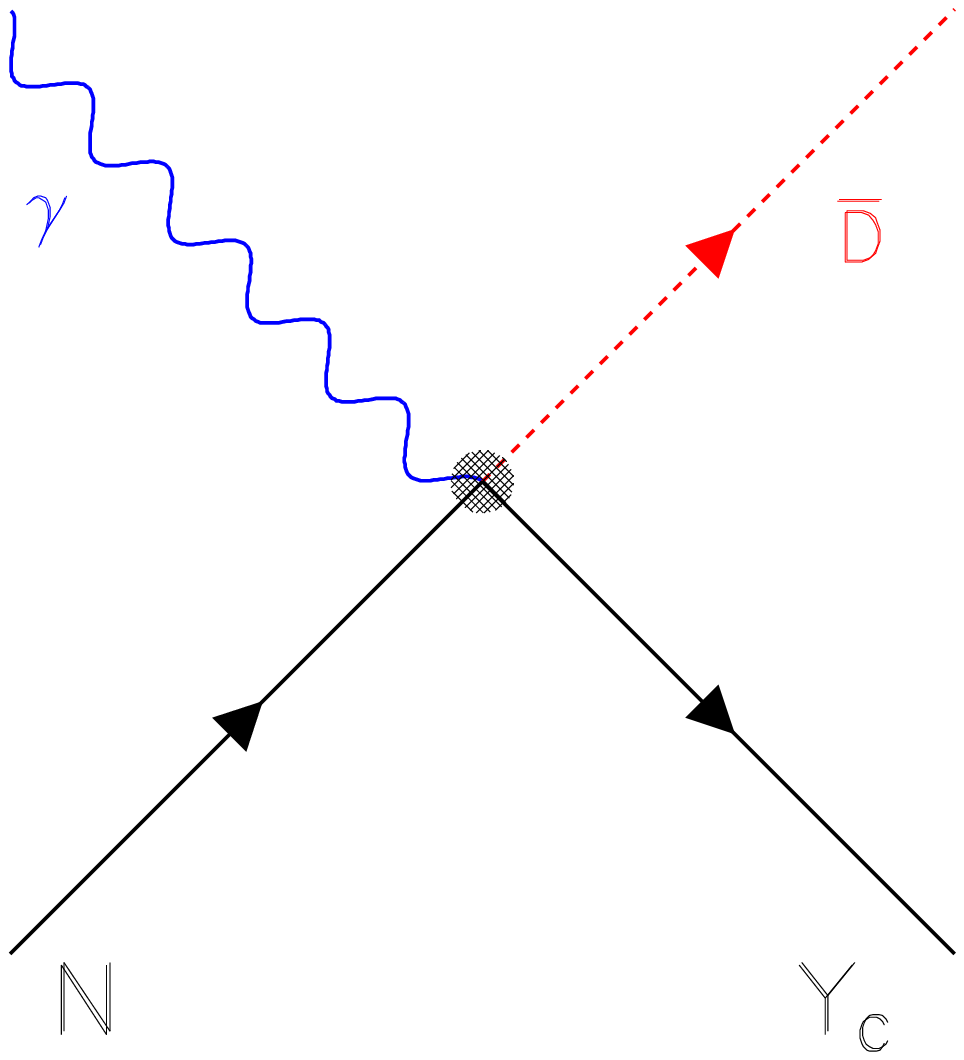}
\caption{\label{fig:fig3} The catastrophic diagram for $\gamma+N\to \overline{D}+Y_c$ for the pseudovector $NY_c\overline{D}$-interaction.}
\end{center}
\end{figure}

Note that only the sum of $s$, $t$, and $u$ exchanges gives a gauge invariant total matrix element for $\gamma+N\to Y_c+\overline{D}$. More exactly this holds for pseudoscalar interaction in the vertex $N\to Y_c+\overline{D_c}$. In case of pseudovector vertex, which can also be considered, the gauge invariance is insured by additional contribution of the so-called 'catastrophic' (contact) diagram, (Fig. \ref{fig:fig3}), which has a definite spin structure and a known coupling constant, $g_{NY_c\overline{D}}$\footnote{Note, in this respect, that the direct estimation of this contact contribution to the cross section of the process $\gamma+p \to \Sigma_c^{++}+D^-_c$ \cite{Ru78} results in a too large cross section in the near threshold region, in contradiction with the experimental data \cite{Ab84}.}.

Following the equivalence theorem, both these approaches result in the same matrix element for the 'electric' interaction, induced by the electric charges of the participating hadrons. But the magnetic moments of the nucleons and the $Y_c$ hyperons produce different results for pseudoscalar and pseudovector interactions, showing the relevance of off-mass shell effects in the considered model.
Of course, the magnetic moments of baryons can not violate the gauge invariance for any numerical value and in any kinematical condition.

To take into account the virtuality of the exchanged hadrons, in this approach, form factors (FFs) are introduced in the pole diagrams. For baryonic exchange the corresponding FFs can be parametrized as \cite{Ha98}:
\begin{equation}
F_N(s)=\displaystyle\frac{\Lambda_N^4}{\Lambda_N^4+(s-m^2)^2},~
F_Y(u)=\displaystyle\frac{\Lambda_Y^4}{\Lambda_Y^4+(u-M^2)^2},~
\label{eq:ff1}
\end{equation} 
where $\Lambda_N$ and  $\Lambda_Y$ are the corresponding cut-off parameters, $s=(k+p_1)^2$, $u=(k-p_2)^2$, $k$, $p_1$ and $p_2$ are the four-momenta of the photon, the nucleon, and the hyperon, respectively, $m$ $(M)$ is the nucleon (hyperon) mass.

For mesonic exchanges, another expression of FFs is taken:
\begin{equation}
F_D(t)=\displaystyle\frac{\Lambda_{1\gamma}^2-m_D^2}
{\Lambda_{1\gamma}^2-t}
\displaystyle\frac{\Lambda_1^2-m_D^2}{\Lambda_1^2-t},~
F_{D^*}(t)=\displaystyle\frac{\Lambda_{2\gamma}^2-m_{D^*}^2}
{\Lambda_{2\gamma}^2-t}
\displaystyle\frac{\Lambda_2^2-m_{D^*}^2}{\Lambda_2^2-t},~
\label{eq:ff2}
\end{equation} 
where $t=(k-q)^2=(p_1-p_2)^2$, $m_D$ $(m_D^*)$ is the mass of $D$ $(D^*)$ meson,
$\Lambda_{1,2\gamma}$ are the cutoff parameters for the electromagnetic vertex, $\gamma +D_V(D_V^*)\to D$, with virtual $D(D^*)$ meson and $\Lambda_{1,2}$ are the cutoff parameters for the strong vertex $N\to Y_c+D_V(D_V^*)$. All these form factors are normalized such that $F_N(s=m^2)=F_Y(u=M^2)=F_D(t=m_D^2)=F_{D^*}(t=m^2_{D^*})=1$.

Only for the magnetic contributions these expressions of FFs do not destroy the gauge invariance of the total matrix elements. Once these FFS, which are different for different diagrams, are introduced,  the other contributions, induced by the electric charges of the particles, will be rearranged in such a way that the gauge invariance is strongly violated. The simplest way to restore the gauge invariance is to  multiply the complete matrix element (for $s+t+u$-contributions) by a common factor \cite{Ha98}:
\begin{eqnarray*}
&\displaystyle\frac{1}{3}\left [F_N(s)+F_Y(u)+F_D(t)\right ], &\mbox{~for~} \gamma+N\to Y_c+D^-, \\
&\displaystyle\frac{1}{2}\left [F_N(s)+F_Y(u)\right ], &\mbox{~for~} \gamma+N\to Y_c+\overline{D^0}. 
\end{eqnarray*}
Such  term is a function of both independent kinematical variables, therefore it can not be rigorously called a form factor, which must be, in general, function of one variable only.

These terms decrease essentially the differential cross section, at large values of $|t|$ or $|u|$, and therefore the total cross section, especially in the near threshold region. The role of FFs is essential for such approach, as it has been proved in the analysis  of vector meson or strange particle production in $NN$-- and $\Delta N$--collisions \cite{Re03a}. Particularly large effects appear for the processes of open charm production in exclusive reactions, such as $N+N\to Y_c+\overline{D}+N$ \cite{Ga02,Re03b}. But polarization phenomena are, in principle, less sensitive to FFs. Moreover, in the limiting case of $s+u+t(D)$ contributions (without $D^*$) or only vector $D^*$-exchange, one can see that polarization observables are independent on any phenomenological FFs, for any kinematics. Such models can generate only $T-$even polarization observables, such as $\Sigma_B$-asymmetry, induced by linearly polarized photons, or $A_{x,z}$-asymmetries, induced by the collision of circularly polarized photons with a polarized proton target.

\section{The matrix elements for  $\gamma+N\to Y_c+\overline{D_c}$, $ Y_c=\Lambda_c,\Sigma_c$}

Let us consider in detail the matrix elements for the reactions of  $\gamma+N\to Y_c+\overline{D_c}$ $ Y_c=\Lambda_c,\Sigma_c$, considering $s$, $u$ and 
$t$($\overline{D}+\overline{D^*}$) contributions:
$$
{\cal M}={\cal M}_s+{\cal M}_u+{\cal M}_t(D)+{\cal M}_t(D^*).
$$
For the pseudoscalar $N\overline{D}_c Y_c$ vertex, the matrix element for one-nucleon exchange in $s$-channel, ${\cal M}_s$, can be written as follows:
\begin{equation}
{\cal M}_s=\displaystyle\frac{eg_{NY_cD}}{s-m^2}\overline{u}(p_2)
\gamma_5 \left (\hat{p_1}+\hat{k}+m \right )
\left ( Q_{N}\hat{\epsilon}-\kappa_N
\displaystyle\frac{\hat{\epsilon}\hat{k}}{2m}\right )u(p_1),
\label{eq:ms}
\end{equation}
where $\kappa_N$ is the anomalous magnetic moment of the nucleon, $g_{NY_cD}$ is the coupling constant for the vertex $N\to Y_c+D$, $Q_N$ is the electric charge of the nucleon. We assume, in Eq. (\ref{eq:ms}) (and later on in this paper), that the relative ${\cal P}$-parity of the $N\overline{D}Y_c$ system is negative, in agreement with the quark model. In principle, this ${\cal P}$-parity can be experimentally determined \cite{Re03c}.

The other matrix elements can be written as:

\begin{eqnarray}
{\cal M}_u&=&\displaystyle\frac{eg_{NY_cD}}{u-m^2}\overline{u}(p_2)\left ( Q_{Y_c}\hat{\epsilon}-\kappa_{Y_c}\displaystyle\frac{\hat{\epsilon}\hat{k}}{2M}\right )
\left (\hat{p_2}-\hat{k}+M \right )\gamma_5 u(p_1), \\
{\cal M}_t(D)&=&\displaystyle\frac{eg_{NY_cD}}{t-m_D^2}Q_{D}\overline{u}(p_2)
\gamma_5 u(p_1)2\epsilon\cdot q,\label{eq:mm}\\
{\cal M}_t(D^*)&=&i\displaystyle\frac{eg_{NY_cD^*}}{t-m_{D^*}^2}
\displaystyle\frac{g_{D^*D\gamma}}{m_{D^*}}\epsilon_{\mu\nu\alpha\beta}
\epsilon_{\mu}k_{\nu}{\cal J}_{\alpha}(k-q)_{\beta},\nonumber\\
\end{eqnarray}
where $Q_{Y_c}$ and  $Q_{\overline{D}}$ are the electric charges of the $Y_c$-hyperon and of the $\overline{D}$ meson, so that $Q_N=Q_{Y_c}+Q_{\overline{D}}$,  $\epsilon_{\nu}$ is the four-vector of photon polarization, ${\cal J}_{\alpha}$ is the vector current for the vertex $N\to Y_c+D^*$:
\begin{equation}
{\cal J}_{\alpha}=\overline{u}(p_2)\left [\gamma_{\alpha}(1+\kappa_{Y_c})-\kappa_{Y_c}
\displaystyle\frac{p_{1\alpha}+p_{2\alpha}}{m+M}\right ]u(p_1),
\end{equation}
where $g_{NY_cD^*}$ and $\kappa_{Y_c}g_{NY_cD^*}$ are the vector (Dirac) and the tensor (Pauli) coupling constants for the vertex $D^*N\to Y_c$ and $g_{D^*D\gamma}$ is the  coupling constant for the vertex $D^*\to D\gamma $. The corresponding width $\Gamma(D^*\to D\gamma)$ in terms of $g_{D^*D\gamma}$ can be written as follows:
$$\Gamma(D^*\to D\gamma) =  \displaystyle\frac{\alpha}{24}m_{D^*}\left (
1- \displaystyle\frac{m_D^2}{m^2_{D^*}}
\right )^3g^2_{D^*D\gamma},~\alpha=\displaystyle\frac{e^2}{4\pi}\simeq\displaystyle\frac{1}{137}.
$$
From the experimental data about $D^{*+}$ decays \cite{PDG}:
$$\Gamma(D^{*+})=(96\pm4\pm 22)~\mbox{keV},~
\mbox{Br}(D^*\to D\gamma)=(1.68 \pm 0.42 \pm 0.49)\%
$$
one can find: $|g_{D^{*+}D^+\gamma}|=1.03$. The sign of this constant can not be determined from these data. The situation with the coupling $g_{D^{*0}D^0\gamma}$ is less definite. Having the largest branching ratio, 
$Br(D^{*0}\to D^0\gamma)\simeq 40$\% , only the upper limit is experimentally known for the total width of neutral  $D^{*0} $: $\Gamma (D^{*0})\le 2.1 $ MeV \cite{PDG}
i.e  $\Gamma (D^{*0}\to D^0\gamma)\le 840 $ keV, i.e. very far from the theoretical predictions \cite{Re01}, with $\Gamma (D^{*0}\to D^0\gamma)\ge 10 $ keV. 

The expressions for the scalar amplitudes $f_i$, corresponding to the different matrix elements (12-15), are given in the Appendix.
\section{Discussion of the results}
The main ingredients of the considered model, which enter in the numerical calculations of the different observables for the exclusive process $\gamma+N\to Y_c+\overline{D}$, are the strong and electromagnetic coupling constants and the phenomenological form factors.

\subsection{The electromagnetic coupling constants}
The electromagnetic characteristics of the charmed particles, as the magnetic moments of the $Y_c$-hyperon and the $g_{D^*D\gamma}$-coupling constants (transition magnetic moments) are not well known. Only the width of the radiative decay $D^{*+}\to D^++\gamma$ (which allows to derive the corresponding coupling constant $g_{D^{*+}D^+\gamma}$) has been directly measured, but not its sign. The magnetic moments of the charmed hyperons and the transition magnetic moment $g_{D^{*0}D^0\gamma}$ are not experimentally known.

Based on the previous experience with the theoretical description of the magnetic moments of strange hyperons and of the transition magnetic moments of $V\to P+\gamma$ (with $V=\rho$, $\omega$, $\phi$,  $K^*$ vector and $P=\pi$, $\eta$, $K$ pseudoscalar light mesons), one can assume that predictions from symmetry considerations may work in this region of hadron electromagnetic interaction.

We will use different theoretical approach to extrapolate these quantities to charm particle electrodynamics, in particular to the magnetic moments of $Y_c$ and to the amplitude of the radiative decay $D^*\to D+\gamma$. Quark models, $SU(4)$-symmetry, QCD dispersion sum rules and effective chiral theories with heavy quarks can also give useful guidelines.

The dependence on the magnetic moments of the charmed baryons has been studied in \cite{Re2}. Here, for all the calculations, we take the $\Lambda_c^+$ values from \cite{Sa94}, but for the $\Sigma_c$-hyperons we take the $U(4)$-predictions \cite{Re1}: 
$$\mu (\Sigma_c^{++})=\displaystyle\frac{2}{3}\mu _p,~\mu (\Sigma_c^{+})=0,~\mu (\Sigma_c^{-})=-\displaystyle\frac{2}{3}\mu _p.
$$

\subsection{Strong coupling constants}

We call strong coupling constants those which involve one nucleonic vertex, $g_{NY_c\overline{D}}$, $g_{NY_c\overline{D^*}}$, and $\kappa_{Y_c}$, $Y_c=\Lambda_c$ or $\Sigma_c$. Six coupling constants enter in the calculation of the different observables (three for $\gamma+N \to \Lambda_c + \overline{D}$ and three for $\gamma+N \to \Sigma_c + \overline{D}$) and of their $E_\gamma$ and $\cos\vartheta$ dependences, for all the possible exclusive reactions of associative charm particle photoproduction, $\gamma+N\to Y_c+\overline{D}$.

Note that the same coupling constants enter in the description of charmed particle production in $\pi N$-- collisions: $\pi+N\to Y_c+ \overline{D}$, $NN$--collisions: $N+ N\to N+Y_c+ \overline{D}$ and in the interaction of charmed particles with nuclei in heavy ion collisions, $D+N\to Y_c+P(V)$, 
$Y_c+N\to N+N+D$ etc. 

The lack of experimental data about these processes do not allow to fix these coupling constants. Therefore the typical way to estimate these couplings is to rely on $SU(4)$ symmetry, and connect the necessary coupling constants with the corresponding constants for strange particle production:
\begin{equation}
g_{N\Lambda (\Sigma)K},~g_{N\Lambda (\Sigma)K^*},~
\kappa_{N\Lambda (\Sigma)K^*},
\label{eq:strp}
\end{equation}
taking into account that the strange quark is the heaviest of the three light quarks $(u,d,s)$, and the charmed quark is the lightest of the three heavy quarks  $(c,b,t)$.

The coupling constants (\ref{eq:strp}) for strange particles, have been estimated from several experiments in photo- and electroproduction  of strange particle on nucleons, $\gamma+N\to \Lambda(\Sigma)+K$ and $e^-+N\to e^- +\Lambda(\Sigma)+K$ \cite{Gu97}.  However, different models predict different sets of  constants (\ref{eq:strp}), which can take values in a wide interval. 

This gives, nevertheless, a starting point of our analysis, applying $SU(4)$ symmetry. We have also to keep in mind that $SU(4)$ symmetry can be strongly violated, at least at the scale of difference in masses of charmed and strange particles (induced by the difference in masses of $c$ and $s$ quarks).

\subsection{Phenomenological form factors}

In order to determine the form factors, one has to choose a convenient analytical parametrization and then numerical values for the cut-off parameters: $\Lambda_N$, $\Lambda_{1,2}$, and $\Lambda_{1,2\gamma}$. Generally one takes a monopole, dipole or exponential dependence on the momentum transfer. One possibility is to choose as argument of these form factors the four-momentum transfer squared, but one can also take the three-momentum transfer, as well. In this last case, the corresponding reference frame has to be indicated. 

The numerical values of the cut-off parameters can be determined from the previous experience in the interpretation of other photoproduction and hadroproduction processes, in framework of the ELA approach. In the present calculations we use the parametrizations for the corresponding form factors, following Eqs.  (\ref{eq:ff1}) and (\ref{eq:ff2}).

\subsection{Three possible scenarios}
Summarizing the previous discussion, one can conclude that $SU(4)$ symmetry gives some guidelines to fix the necessary parameters of the calculation (strong coupling constants and $Y_c$ magnetic moments). Note that not only the absolute values of these couplings are important in our considerations, but also their relative signs, due to the strong interference effects between the different contributions. The relative signs of $s$-, $u$- and $t(D)$-contributions to the matrix element of any exclusive process $\gamma +N\to Y_c+\overline{D}$ are uniquely fixed by the requirement of gauge invariance. But it is not the case for the relative sign of $D^*$ contribution, so one must do some assumptions, as the validity of $SU(6)$-symmetry. It is important to stress that such symmetry consideration is very powerful for the prediction of relative signs, in contrast with the predictions of the absolute values for the considered couplings.

Any fitting procedure induces a strong correlation between the values of the cut-off parameters and of the strong coupling constants. Therefore there is no unique solution, and we will consider three possible scenarios.

\begin{enumerate}
\item The strong coupling constants are fixed by $SU(4)$-symmetry, using the corresponding values for the strange coupling constants (\ref{eq:strp}), found in the analysis of experimental data \cite{Gu97} on associative strange particles photo- and electroproduction. Only the cut-off parameters are fixed on the charm photoproduction data. We assume for simplicity:
$$ \Lambda_1=\Lambda_2=\Lambda_{1\gamma}=\Lambda_{2\gamma}\equiv \Lambda,~\Lambda_Y=\Lambda_N. $$

\item We assume that $SU(4)$-symmetry is strongly violated for the strong coupling constants. So we take for the couplings (\ref{eq:strp}) some arbitrary values, far from $SU(4)$-predictions, and both cut-off parameters, $\Lambda$ and $\Lambda_N$, are determined from charm photoproduction data.

\item We assume that $SU(4)$-symmetry is violated only for the $N\to \Lambda_c+\overline{D}^*$-vertex, but for the couplings $g_1$ and $g_2$ we take the $SU(6)$ values of the corresponding coupling constants for the vertex $N\to \Lambda+K^*$. Again the parameters $\Lambda_N$ and $\Lambda$ are fixed from charm photoproduction data at $E_{\gamma}$=20 GeV. 
\end{enumerate}

Taking into account the limited experimental data about charm particles photoproduction, one can not do a rigorous fit for all the parameters of the model.

The model can predict the energy behavior of the total cross section for $\gamma^*+N\to Y_c+D$ (for proton and neutron targets), for each of the three versions. Therefore we normalize the total cross section for $\gamma +p\to \Lambda_c^+ +\overline{D^0}$, which is the largest from all the exclusive reactions  $\gamma +p\to Y_c +\overline{D}$, to the measured cross section of open charm photoproduction at $E_{\gamma}=20$ GeV \cite{Ab86}, where it was found that about 70 \% of the total cross section can be attributed to  $\gamma +p\to \Lambda_c^+ +\overline{D^0}$. This condition constrains very strongly the parameters for all the three versions of the model, where this reaction has the largest cross section.

The parameters are reported in Table \ref{tab:tab1}, where we used the following notations: $g_1=g_{N\Lambda_c D^*}$,~ $g_2=g_{N\Lambda_c D^*}\kappa_{\Lambda_c}$.
 
\begin{table}[h]
\begin{tabular}{|c|c|c|c|c|c|}
\hline\hline
Model &$g_{N\Lambda_cD}$ &$g_1$&$g_2$&$\Lambda_N$& $\Lambda $\\
\hline\hline
I&-11.5& -23& -57.5 & 0.8 & 2.4\\
\hline
II&-2.& -2.5& -6. & 1.6 & 3.4\\
\hline
III &-2.&  -6.&  -22. &  0.6 &  2.7\\
\hline\hline
\end{tabular}
\caption{ Parameters for models I, II, III, see text. The cut-off parameters $\Lambda_N$ and $\Lambda$ are expressed in [GeV].}
\label{tab:tab1}
\end{table}

Our procedure is not a real fit, as we take into account only the low energy point, therefore we do not give any $\chi^2$ estimates of the quality of the model, in its three versions.

Once the parameters have been fixed, Table (\ref{tab:tab1}), one can predict all polarization observables not only for  $\gamma +p\to \Lambda_c^+ +\overline{D^0}$  but also for any exclusive reaction $\gamma +N\to Y_c^+ +\overline{D}$. For the reactions with $\Sigma_c$-production, we take the following coupling constants: $g_{N\Sigma_c \overline{D}}$=4.5, $g_1(N\Sigma_c \overline{D^*})$=-$g_2(N\Sigma_c \overline{D^*})$=-25, which correspond to the values of  $g_{N\Sigma K}$ and  $g_{N\Sigma K^*}$, obtained form a fit to the experimental data about $\gamma+p\to \Sigma^0+K^+$ and $e+p\to e+\Sigma^0+K^+$ \cite{Gu97}. For the cutoff parameters we took the values $\Lambda_N=0.8$ and $\Lambda$=2.4 GeV, as for model I.

\begin{figure}
\includegraphics[width=15cm]{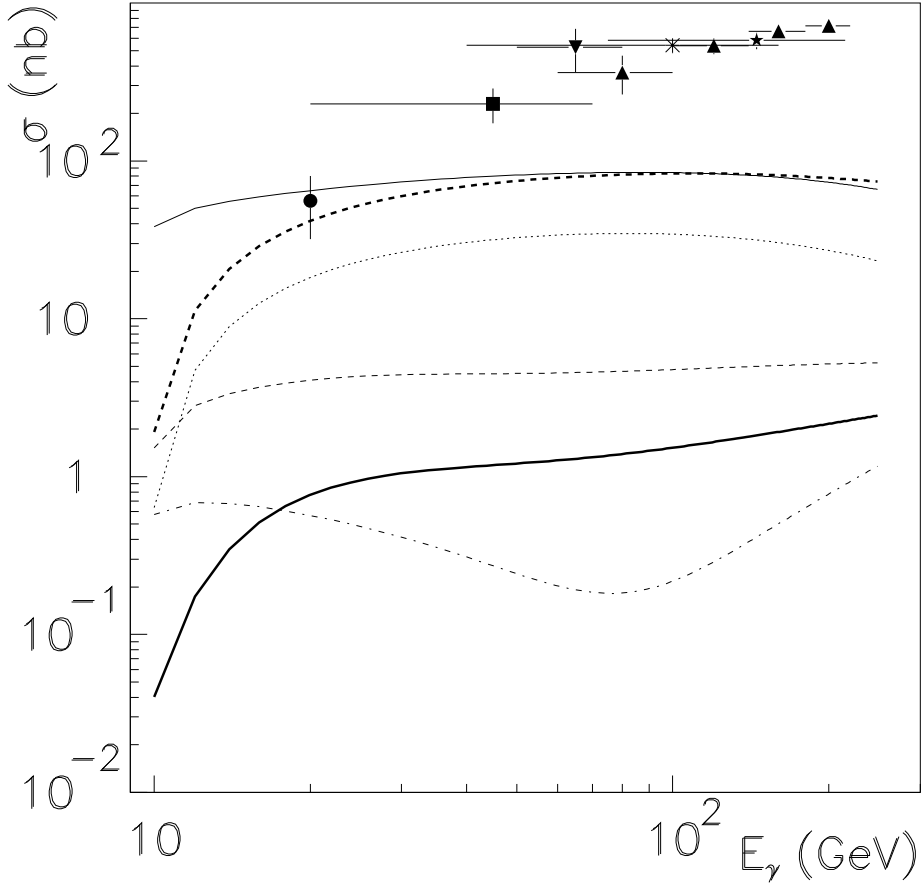}
\caption{\label{fig:exc0} $E_\gamma$-dependence of the total cross section for photoproduction of charmed particles for model I. The curves correspond to different reactions: $\gamma+p\to \Lambda_c^+ 
+\overline{D^0}$ (solid line),
$\gamma+p\to\Sigma_c^{++}+D^-$ (dashed line), 
$\gamma+p\to \Sigma_c^+  + \overline{D^0}$ (dotted line), 
$\gamma+n \to \Lambda_c^+ +D^-$  (dot-dashed line), 
$\gamma+n \to \Sigma_c^+  + D^-$ (thick solid line), 
$\gamma+n \to \Sigma_c^0  + \overline{D^0}$ (thick dashed line). The data correspond to the total charm photoproduction cross section from \protect\cite{As80} (reverse triangle), \protect\cite{Ad87} (square), \protect\cite{Al92} (asterisk),  \protect\cite{Anjos89} (star),
\protect\cite{Au83} (triangles), \protect\cite{Ab86} (circle), }
\end{figure}

The energy dependence of the total cross section for the six exclusive processes $\gamma+N\to Y_c+\overline{D}$, is very different, in all the photon energy range, Fig. \ref{fig:exc0}. For $E_{\gamma}\ge 40$ GeV, the predicted energy dependence of the total cross section becomes flat, up to  $E_{\gamma}= 250$ GeV. One can conclude that, in this energy range,  the simplest exclusive photoproduction reactions $\gamma +N\to Y_c^+ +\overline{D}$ contribute less than 10 \% to the total cross section of open charm  photoproduction. This estimation does not contradict the existing experimental data and is in agreement with the measured $\Lambda/\Lambda_c$ asymmetry (in sign and value).

\begin{figure}
\includegraphics[width=15cm]{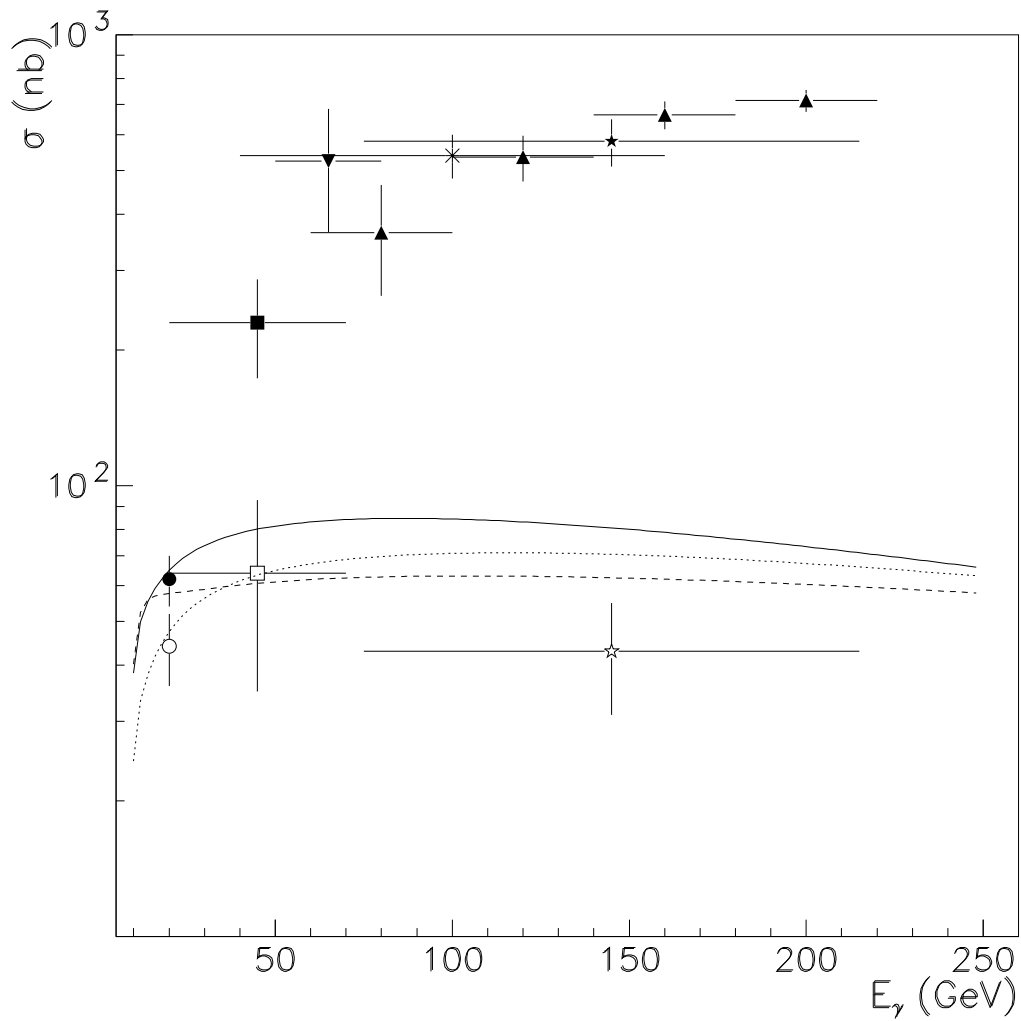}
\caption{\label{fig:exc1}$E_\gamma$-dependence of the total cross section for the 
reaction $\gamma +p\to \Lambda_c^++\overline{D^0}$ for  for model I (solid line), II (dashed line) and III (dotted line). The experimental points are from  \protect\cite{Ab86} (open circle), from \protect\cite{Ad87} (open square), and from  \protect\cite{Anjos89} (open star).
}
\end{figure}

It is interesting to note that we have very large isotopic effects, i.e. a large difference in the absolute values and behavior for the different channels, with different charges of the participating hadrons. This is an expected property of ELA approach, because the relative values of $s$, $u$, and $t$-channel contributions are different for the different channels. Note that the largest cross section on the neutron target belongs to the process  $\gamma +n\to \Sigma_c^0 +\overline{D^0}$, the $D^-$ production being essentially suppressed. Moreover, 
the $D^-$ production is also small in the $\gamma p$ interaction,  $\gamma +p\to \Sigma_c^{++} +{D^-}$, in agreement with the experiment \cite{Ad87}.
The total cross section as a function of the photon energy, 
for the reaction $\gamma+p\to \Lambda_c^++\overline{D^0}$, is shown in Fig. \ref{fig:exc1}, for the three sets of parameters.

\subsection{Contribution to $\overline{D}/{D}$ asymmetries }

Our aim here is to have a realistic view on general characteristics of the different reaction channels, for charm photoproduction, in a region which is accessible by experiments. A large antiparticle/particle asymmetry, not explainable in terms of pQCD models, has been reported in the literature and numerical  estimations of associative charm photoproduction cross sections have  been done in Refs. \cite{Ab86,Ad87,Anjos89}. The values are reported as open symbols in Fig. \ref{fig:exc1}.

The observation of the $\overline{D^0}/D^0$ or $\overline{\Lambda_c}/\Lambda_c$ asymmetry in $\gamma N$-collisions is important in order to test the validity of the photon-gluon fusion mechanism. The discussed asymmetries are defined as:
$$ A=\displaystyle\frac{N(\overline{c})-N(c)}{N(\overline{c})+N(c)},$$ where $N(c)$ [$N(\overline{c})$] is the number of corresponding charm particles ($D$ or $\Lambda_c$) containing $c(\overline{c})$-quarks.

Note that the exclusive photoproduction of open charm in the processes $\gamma+N\to Y_c+\overline{D}$ $\overline{D}^*$ will result in $\overline{D^0}/D^0$ and $\overline{\Lambda_c}/\Lambda_c$ asymmetry in $\gamma N$-collisions (with unpolarized particles), increasing the  $\overline{D}$-production and decrasing the $\overline{\Lambda_c}$-production. Such asymmetries have been experimentally observed \cite{Bi99}. For example, the FOCUS experiment found an asymmetry for $\overline{\Lambda_c}/\Lambda_c$ production of $\simeq -(0.14\pm 0.02)$ at $E_{\gamma}\simeq 180$ GeV, demonstrating that $\Lambda_c^+$-production is more probable than $\overline{\Lambda_c}$ \cite{Focus}.  At similar energies the E687 experiment finds enhancement of $\overline{D}$ over ${D}$ production \cite{Fr96}. Very large (in absolute value) charm asymmetries have been observed also at $E_{\gamma}$=20 GeV, at SLAC \cite{Ab86}. Note also that the SELEX collaboration presented results on the $\overline{\Lambda_c}/\Lambda_c$--asymmetries for different hadronic processes: $p$, $\pi^-$, $\Sigma^-+N\to$ $\Lambda_c^{\pm} +X$ \cite{Se01}. Similar results have been presented by the Fermilab E791 collaboration \cite{Ai00}. This is in contradiction with the model of photon-gluon fusion, which predicts symmetric $\overline{\Lambda_c}-\Lambda_c$ yields and can be considered as an indication of the presence of other mechanisms. The exclusive processes, $\gamma+N\to Y_c+\overline{D_c}$, discussed in this paper, explain naturally such asymmetry.

\subsection{Differential cross section and polarization observables}

\begin{figure}
\includegraphics[width=15cm]{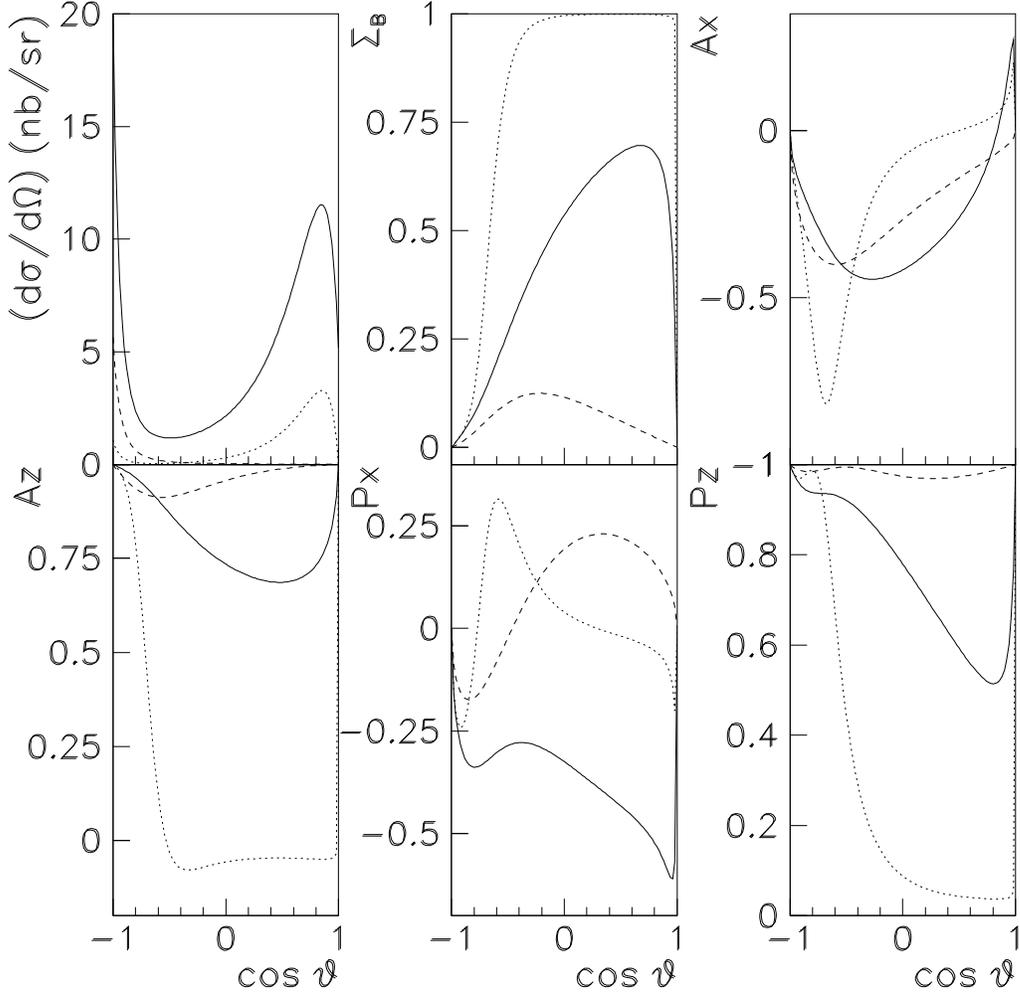}
\caption{\label{fig:angr1a} Differential cross section and polarization observables $\Sigma_B$, $A_x$, $A_z$, $P_x$, and $P_z$, for the reactions: $\gamma +p\to \Lambda_c^++\overline{D^0}$  (solid line), $\gamma +p\to \Sigma_c^{++}+{D^-}$ (dashed line) and $\gamma +p\to \Sigma_c^++\overline{D^0}$ (dotted line), calculated for model I.
}

\end{figure}

\begin{figure}
\includegraphics[width=15cm]{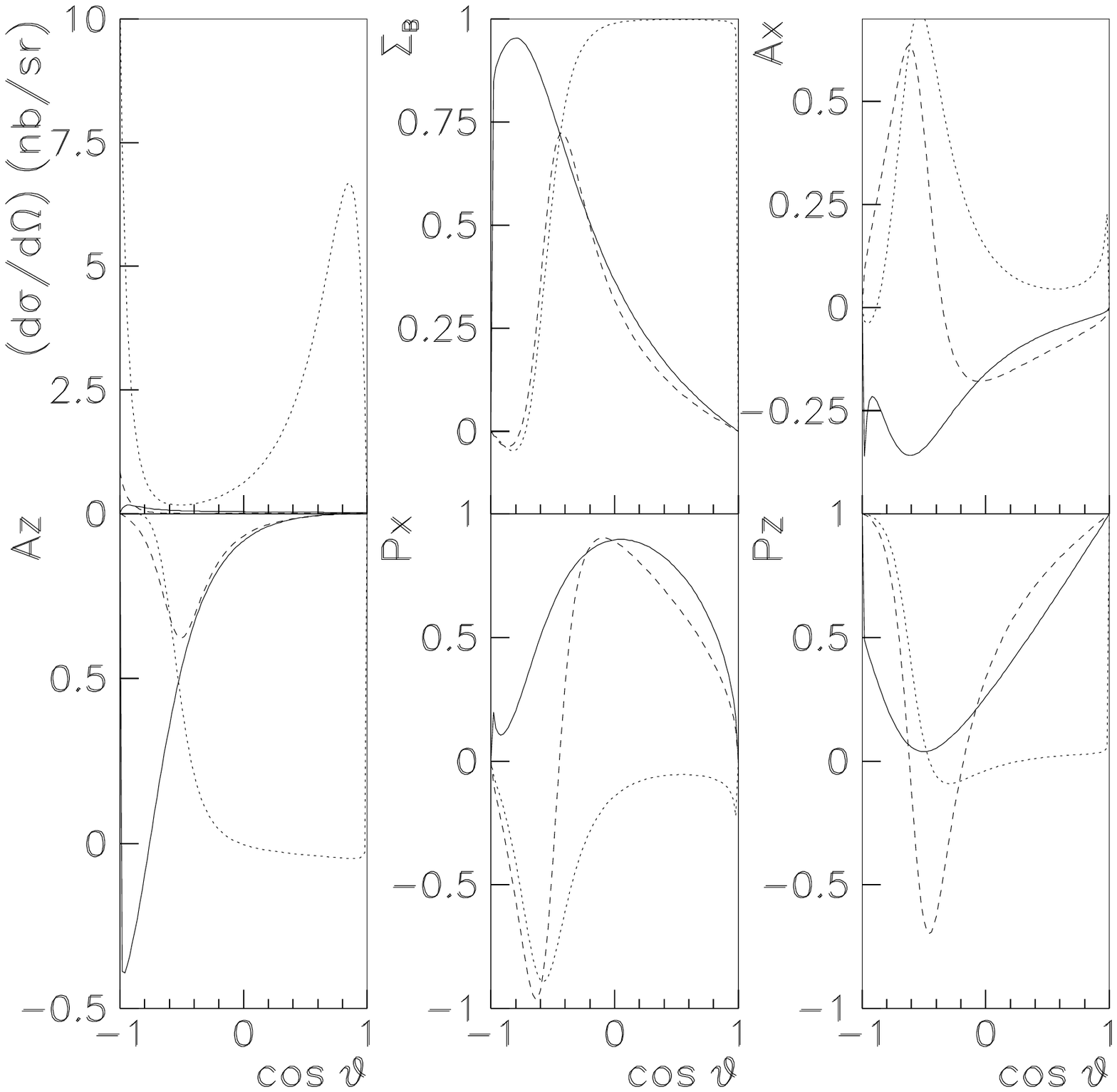}
\caption{\label{fig:angr1c} Differential cross section and polarization observables $\Sigma_B$, $A_x$, $A_z$, $P_x$, and $P_z$, for the reactions: $\gamma +n\to \Lambda_c^++{D^-}$  (solid line), $\gamma +n\to \Sigma_c^++{D^-}$ (dashed line) and $\gamma +n\to \Sigma_c^0+\overline{D^0}$ (dotted line), calculated for model I.
}

\end{figure}

The prediction of the $\cos\vartheta$--dependence of  $(d\sigma/d\Omega)_0$, $\Sigma_B$, $A_x$, $A_z$,  $P_x$, and  $P_z$, for the six processes $\gamma+N\to Y_c+\overline{D}$ $(Y_c=\Lambda_c^+$, $\Sigma_c$
) - on proton and neutron targets, are shown in Figs. \ref{fig:angr1a} and \ref{fig:angr1c}, for model I, at  $E_{\gamma}$=15 GeV. 

Note that, for any version of the considered model, the asymmetry $\Sigma_B$ is positive in the whole angular region, in contradiction with the predictions of PGF \cite{Du80} and in agreement with the SLAC data \cite{Ab86}.

\begin{figure}
\includegraphics[width=15cm]{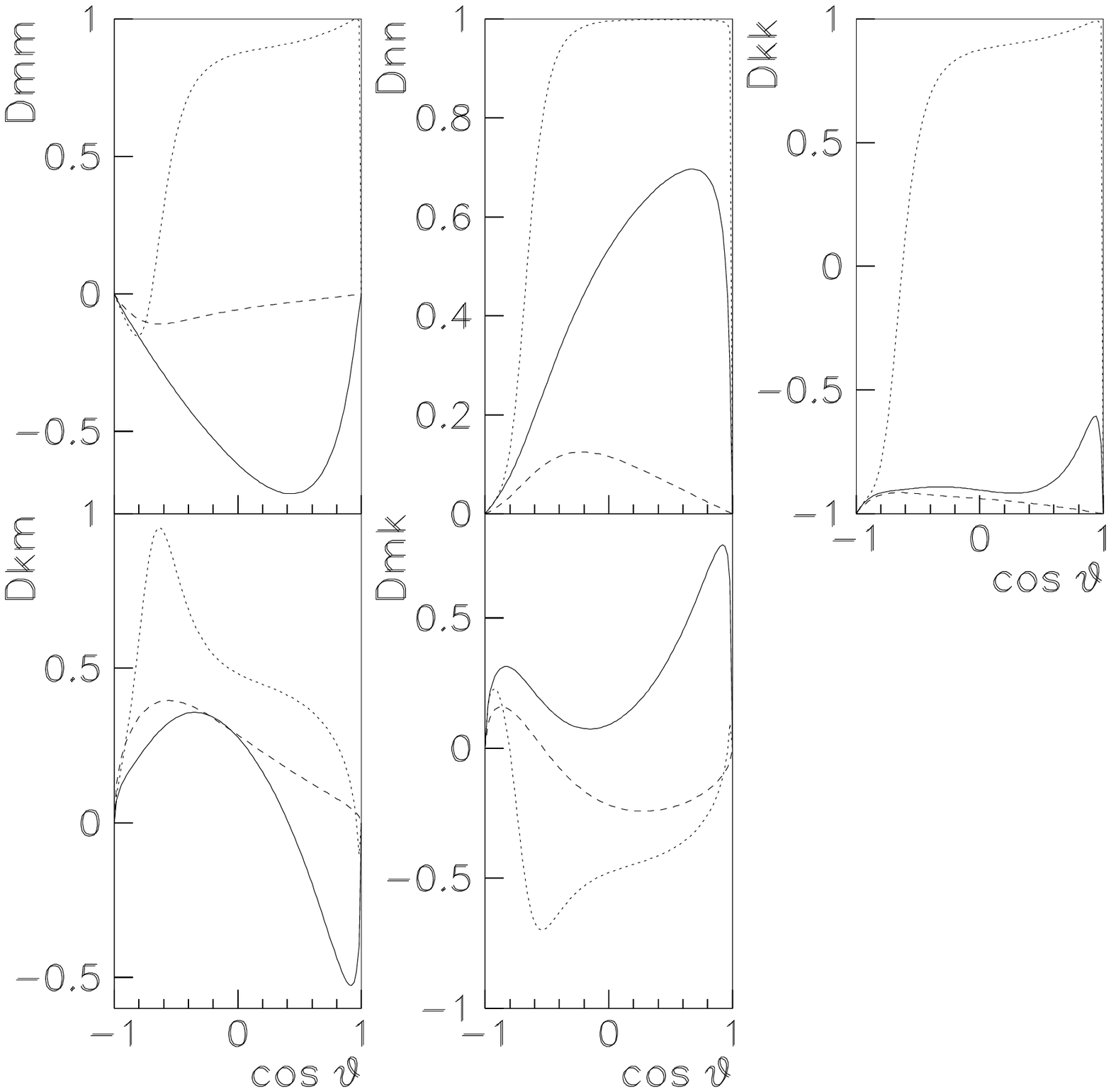}
\caption{\label{fig:angr1b} Depolarization coefficients $D_{ab}$ for the reactions: $\gamma +p\to \Lambda_c^++\overline{D^0}$  (solid line), $\gamma +p\to \Sigma_c^{++}+{D^-}$ (dashed line) and $\gamma +p\to \Sigma_c^++\overline{D^0}$ (dotted line), calculated for model I.
}

\end{figure}

\begin{figure}
\includegraphics[width=15cm]{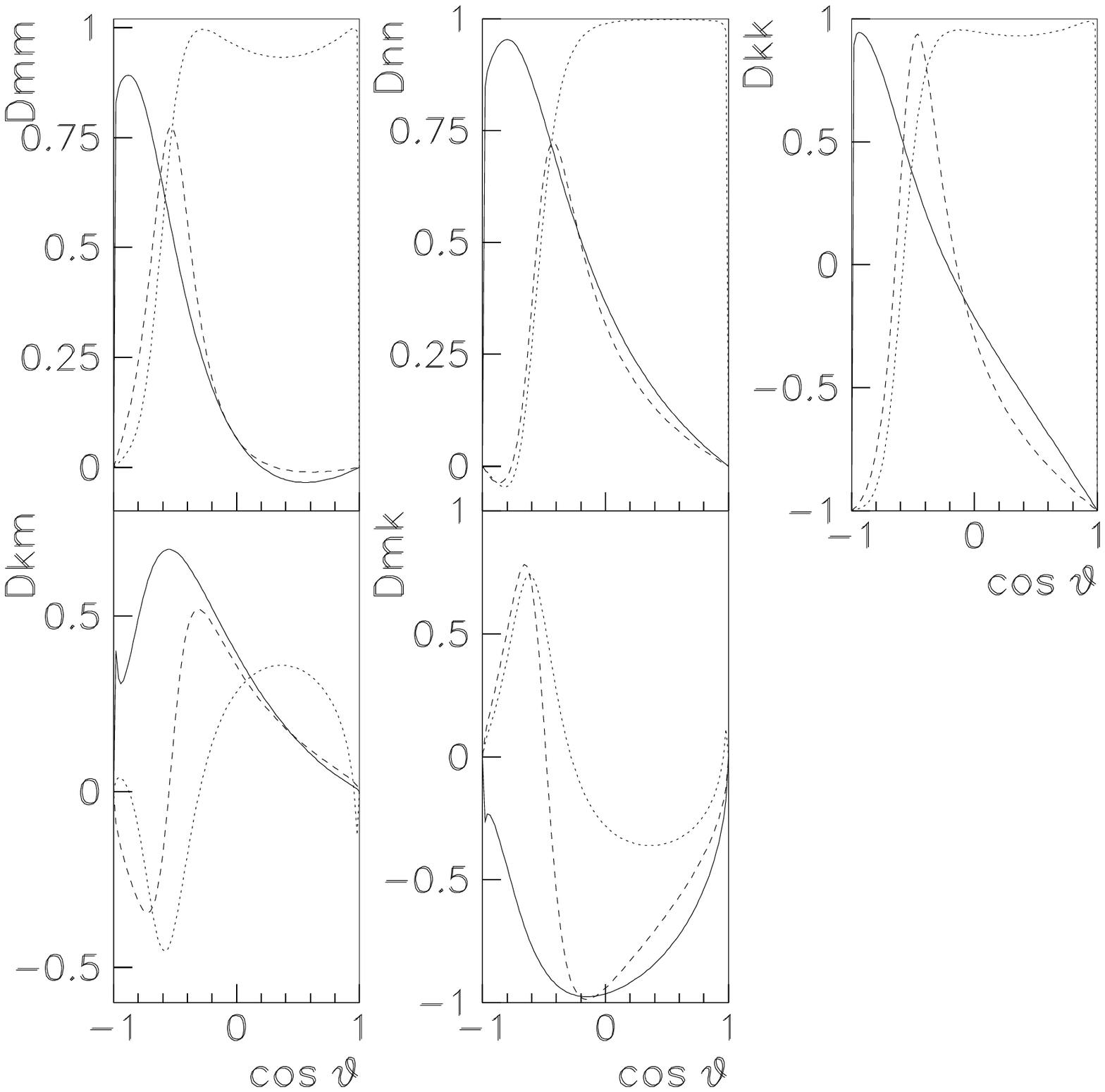}
\caption{\label{fig:angr1d} Depolarization coefficients $D_{ab}$ for the reactions: $\gamma +n\to \Lambda_c^++{D^-}$  (solid line), $\gamma +n\to \Sigma_c^++{D^-}$ (dashed line) and $\gamma +n\to \Sigma_c^0+\overline{D^0}$ (dotted line), calculated for model I.
}

\end{figure}

At the same energy, for the same reactions, with similar notations, the depolarization coefficients $D_{ab}$ are shown in Figs. \ref{fig:angr1b} and \ref{fig:angr1d}.

Polarization effects are generally large (in absolute value), characterized by a strong $\cos\vartheta$-dependence, which results from a coherent effect of all the considered pole contributions. Large isotopic effects (i.e. the dependence on the electric charges of the participating hadrons) are especially visible in the $\cos\vartheta$-distributions for all these observables.


\begin{figure}
\begin{center}
\includegraphics[width=15cm]{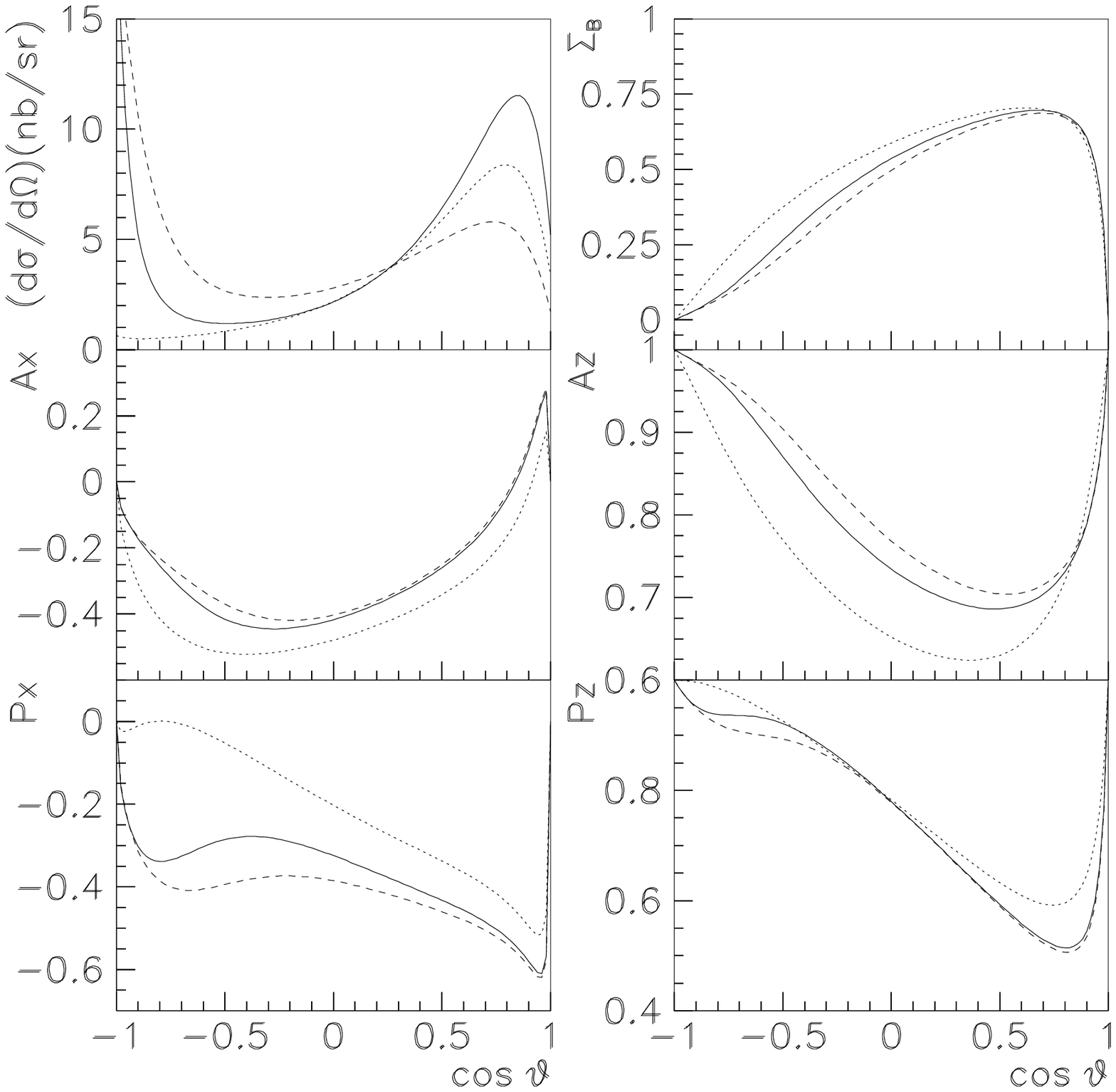}
\caption{\label{fig:mod} $\cos\vartheta$-dependence of the differential cross section, beam asymmetry $\Sigma_B$, and polarization observables: $A_x$, $A_z$,  $P_x$, and  $P_z$, for the 
reaction $\gamma+p\to \Lambda_c^++\overline{D^0}$ for  model I (solid line), II (dashed line) and III (dotted line).
}
\end{center}
\end{figure}

\begin{figure}
\begin{center}
\includegraphics[width=15cm]{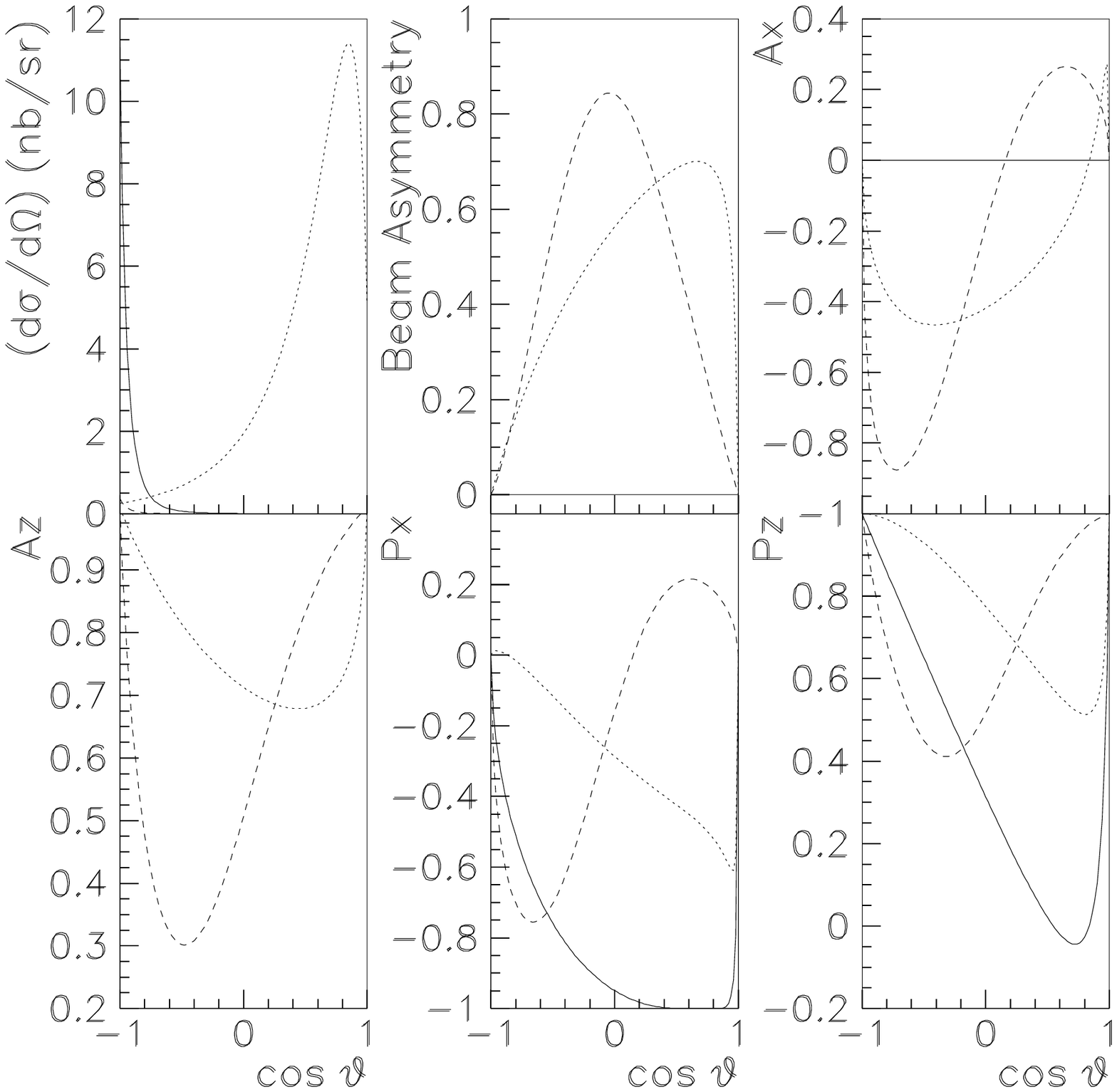}
\caption{ $\cos\vartheta$-dependence of the different contributions to the total amplitude for the differential cross section, the beam asymmetry, $\Sigma_B$,  and the polarization observables: $A_x $, $A_z$, $P_x $ and  $P_z$,  $s$-channel (solid line),  $u$-channel (dashed line), 
$D^*$ $t$-channel (dotted line). The first contribution gives $\Sigma_B$=0 and $A_z$=1, in all kinematical range, because $f_{3,s}=f_{4,s}=0$.}
\label{fig:contr} 
\end{center}
\end{figure}

The dependence of these observables on the version of the model, at $E_{\gamma}$=15 GeV, for the reaction $\gamma +p\to \Lambda_c^++\overline{D^0}$ is shown in Fig. \ref{fig:mod}. For the same reaction, the $\cos\vartheta$-dependence of the individual $s$, $u$ and $t(D^*)$-contributions to the differential cross section and to the considered polarization observables, is shown in Fig. \ref{fig:contr}. This behavior is the same for any version of the considered model, the difference appearing in the interference of the different contributions.

\begin{figure}
\begin{center}
\includegraphics[width=15cm]{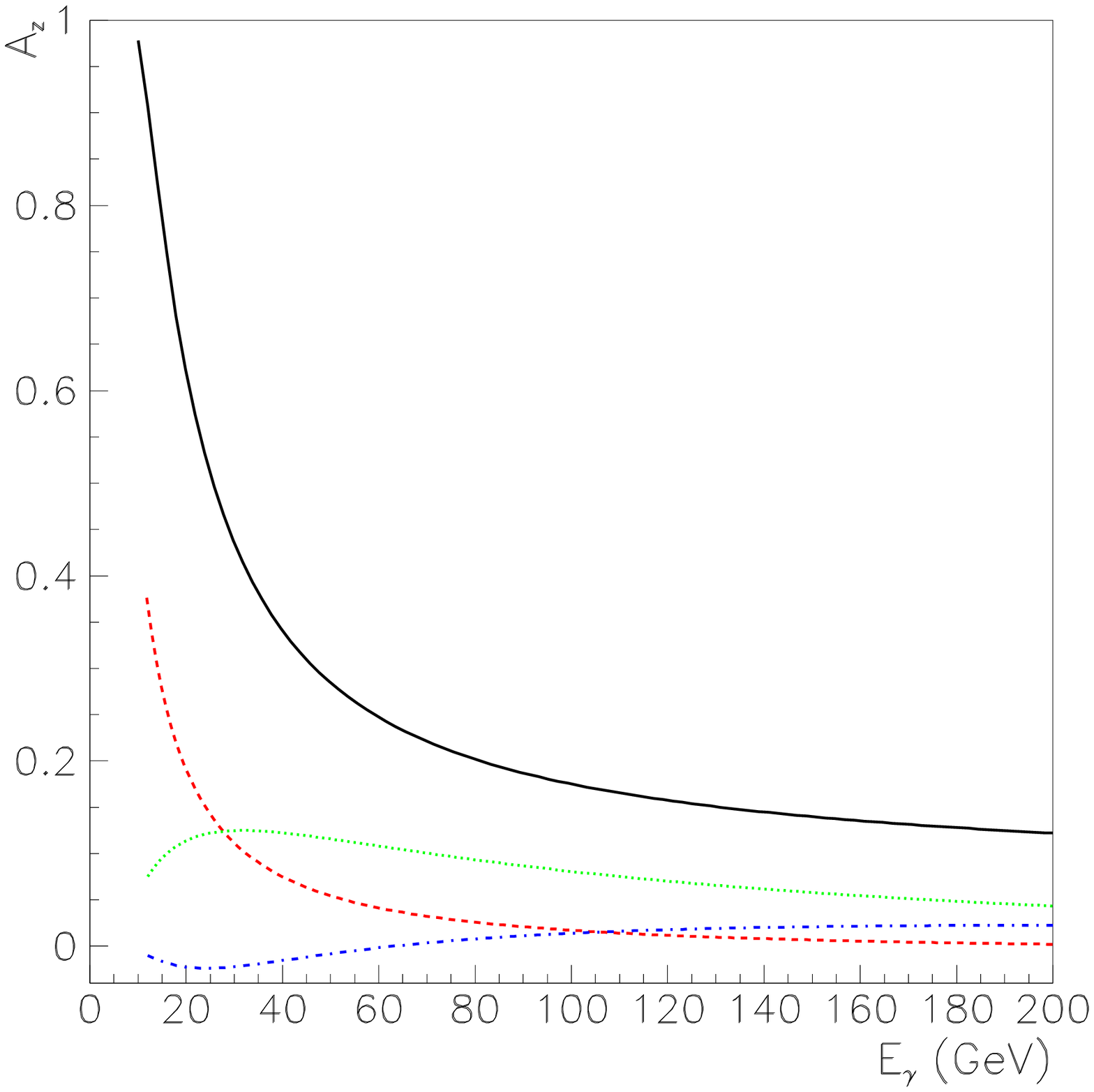}
\caption{Integrated $\overline{A_z}(E_{\gamma})$--asymmetry for the reaction $\vec\gamma +\vec p\to \Lambda_c^++\overline{D^0}$ for ELA approach (solid line), for model I. Predictions for the asymmetry $A_{\gamma N}^{c\overline{c}}(E_{\gamma })$ for inclusive charm photoproduction  calculated with Eq. \ref{eq:eq3}, taking $G(x)$ and $\Delta G(x)$ from  \protect\cite{Br97} (dashed line), from \protect\cite{Gs96} model B (dotted line), and from \protect\cite{Gs96} model C (dashed-dotted line), are also shown.}
\label{fig:azsym} 
\end{center}
\end{figure}

Fig. \ref{fig:azsym}  shows the energy dependence (for model I) of the integrated $\overline{A_z}( E_{\gamma}) $-asymmetry, for $\vec\gamma +\vec p\to \Lambda_c^++\overline{D^0}$, which is defined as:
$$\overline{A_z}(E_{\gamma})=\displaystyle\frac{\int_{-1}^{+1} {\cal N}_0 A_z( E_{\gamma},\cos\vartheta) d\cos\vartheta}{\int_{-1}^{+1} {\cal N}_0d\cos\vartheta}.$$
This asymmetry is large near threshold, taking its maximum value $A_z=+1$ at threshold. It decreases with energy, due to the facts that the cross section increases with energy and then flattens, whereas the $t(D^*)$ contribution becomes more important.  The interference of the $t(D^*)$-channel with  $s$-- and $u$ --channels essentially decreases the $A_z( E_{\gamma})$ asymmetry, outside collinear kinematics, where $A_z( E_{\gamma})=+1$ for all contributions. 

For comparison, predictions for this observable in inclusive charm photoproduction  are also shown. These calculations have been done in framework of standard QCD approach, assuming the PGF model (Fig. \ref{fig:fig1}), doing the ratio of the elementary cross sections for $\gamma+G\to c+\overline{c}$ folded with the gluon distributions:
\begin{equation}
A_{\gamma N}^{c\overline{c}}(E_{\gamma })=\displaystyle\frac{\Delta\sigma_{\gamma N}^{c\overline{c}}}{\sigma_{\gamma N}^{c\overline{c}}}=
\displaystyle\frac{\int_0^1 dx \Delta\sigma(\hat s)\Delta G(x)}
{\int_0^1 dx \sigma(\hat s) G(x)}=
\displaystyle\frac{\int_{4m_c^2}^{2M E_{\gamma}} d\hat s\Delta\sigma(\hat s)\Delta G(x)}
{\int_{4m_c^2}^{2M  E_{\gamma}} d\hat s \sigma(\hat s) G(x)},
\label{eq:eq3}
\end{equation}
where $G(x)$ [$\Delta(G)(x)$] is the unpolarized [polarized] gluon distribution in an unpolarized [polarized] proton and $\hat s$ is the invariant mass of the photon-gluon system and $m_c$ is the charm quark mass.

Several parametrizations exist for the $G(x)$ and $\Delta G(x)$, but, if the unpolarized distribution, $G(x)$, is quite well constrained from the deep inelastic scattering (DIS) data, and, therefore, different calculations give similar results, the polarized gluon distribution $\Delta G$ is poorly known.
For illustration,  $G(x)$ and $\Delta G(x)$
taken from \protect\cite{Br97} (dashed line) and from \protect\cite{Gs96} model B (dotted line) and model C (dashed-dotted line) are shown in Fig. \ref{fig:azsym}.

The predictions of the asymmetry $A_{\gamma N}^{c\overline{c}}(E_{\gamma })$ strongly depend on the choice of $\Delta G(x)$. Moreover, the results are very sensitive to the lower limit of the integral, i.e. to $m_c$. The value of the charm quark mass (so called the current mass) is known from studies of the charmonium properties \cite{PDG},
$m_c=(1.15\div 1.35) $ GeV, but, for the gluon distribution, values of $m_c$ in the range $m_c=(1.5\div 1.7)$ GeV, are more often used. In the calculations we have assumed that the fitting parameters for the functions $G(x)$ and $\Delta G(x)$ have a weak $m_c$-dependence, and we have taken for all calculations $m_c=1.5$ GeV.

Due to the large value of $\overline{A_z}(E_{\gamma})$, we can conclude that the exclusive process of associative particle production, $\gamma +p\to \Lambda_c^++\overline{D^0}$, has to be taken into account as possible background for the extraction of the polarized gluon distribution, $\Delta G(x,Q^2)$ from the measurement of the $A_z$-asymmetry in the inclusive process $\vec\gamma +\vec N\to charm+X$. The importance of exclusive processes $\gamma +p\to Y_c^++\overline{D}$, $\overline{D}^*$ in the estimation of the asymmetry $A_z$ in open charm photoproduction has been mentioned earlier \cite{Ma99,Ry99}.

\section{Conclusions}

We calculated the differential and total cross sections  for the exclusive  processes $\gamma+N\to Y_c+\overline{D_c}$. We also calculated a set of  T- and P-even polarization observables, such as the $\Sigma_B$--asymmetry, induced by 
a linearly polarized photon beam  on an unpolarized nucleon target,
the asymmetries $A_x$ and $A_z$, in the collisions of circularly polarized photons with a polarized nucleon target, in the reaction (i.e, $xz$)-plane,
the $P_x$ and $P_z$ components of the $Y_c$ polarization, in collisions of circularly polarized photons with unpolarized target.

In framework of the effective Lagrangian approach, we suggest a model for these processes on proton and neutron targets. The gauge invariance of the charm particle electromagnetic interaction drove our choice of specific pole diagrams.
The necessary parameters of the model, such as the magnetic moments of the $Y_c$ hyperons and the strong coupling constants (for the vertices $N\to Y_c+\overline{D}$ and  $N\to Y_c+\overline{D}^*$) have been determined from $SU(4)$-symmetry. The phenomenological form factors, which are essential ingredients in this approach, have been taken in such a way to conserve the gauge invariance of the model, for any value of the coupling constants and cut-off parameters, and for any kinematical conditions.

The parameters of the suggested model, in particular the cut-off parameters for the meson and baryon exchanges, were fixed in order to reproduce the value of the total cross section for $\gamma+p\to \Lambda_c^+ + \overline{D}^0$, at $E_{\gamma}$=20 GeV (the smallest energy where open charm photoproduction has been measured).

The existing experimental information does not allow to fix uniquely the parameters. Therefore, we considered three versions of the suggested model, with different sets of coupling constants and cut-off parameters, which all reproduce the cross section at  $E_{\gamma}$=20 GeV.

We predicted the $\cos\vartheta$-dependence of different polarization observables, which are, in principle, accessible now by the running COMPASS experiment, for example.

Large isotopic effects in the energy and $\cos\vartheta$-dependence of all these polarization observables, as well as the large polarization effects (in absolute value) are a general property of the considered model.

The knowledge of the coupling constants and of the cut-off parameters, will be very useful also for any future calculations of electroproduction of charm particles, $e^-+N\to e^-+Y_c+\overline{D}$ and for photo and electroproduction of charmed vector mesons,  $\gamma +N\to \overline{D^*}+Y_c$, $e^- +N\to e^- +\overline{ D^*}+Y_c$. The same constants enter also in the estimation of $Y_c\overline{D}$-associative production in neutrino-nucleon collisions, induced by neutral and weak currents and for the processes $\pi+N\to N+  Y_c+\overline{D}(\overline{D^*})$ and $N+N\to N+Y_c+\overline{D}(\overline{D^*})$. All these different processes can be calculated in framework of ELA approach, in particular in the near threshold region.

Contributing for about 10 \% to the total cross section of open charm photoproduction on nucleons (for $40\le E_{\gamma}\le 250$ GeV), the exclusive process $\gamma+p\to \Lambda_c^+ + \overline{D}^0$ has to be considered an important background in the interpretation of possible polarization effects in the $\gamma +N\to X$ +charm processes. This refers especially to the $A_z$ asymmetry in  $\vec\gamma +\vec N\to X$ +charm, which is considered as the most direct way to measure the gluon contribution to the nucleon spin. 

\section{Acknowledgments}

We thank the members of the Saclay group of the COMPASS collaboration,for interesting discussions and useful comments.

\section{Appendix}

Here we give the expressions for the scalar amplitudes $f_i$, $i=1-4$:
$$ f_i=f_{i,s}+f_{i,u}+f_{i,t}(D)+f_{i,t}(D^*),$$
where the indices $s$, $u$, and $t$ correspond to $s$, $u$, and $t$ channel contributions.
\begin{itemize}
\item $s$-channel:
$$f_{1,s}=g_{NY_cD}\displaystyle\frac{e}{W+m}\left [ 
Q_N-(W-m)\displaystyle\frac{\kappa_N}{2m}\right ],$$
$$f_{2,s}=g_{NY_cD}\displaystyle\frac{e}{W+m}\left [ 
-Q_N-(W+m)\displaystyle\frac{\kappa_N}{2m}\right ]
\displaystyle\frac{|\vec q|}{E_2+M},$$
$$f_{3,s}=f_{4,s}=0,$$
\item  $u$-channel :
$$f_{1,u}=e\displaystyle\frac{g_{NY_cD}}{u-M^2}
\left \{ Q_{Y_c} (W-M)-\displaystyle\frac{\kappa_c}{2M}
\left [ t-m_D^2+(W-m)(W-M)\right ]\right \},$$
$$f_{2,u}=-e\displaystyle\frac{g_{NY_cD}}{u-M^2}
\displaystyle\frac{|\vec q|}{E_2+M}
\left \{ Q_{Y_c} (W+M)+\displaystyle\frac{\kappa_c}{2M}
\left [ t-m_D^2+(W+m)(W+M)\right ]\right \}\displaystyle\frac{W-m}{W+m},$$
$$f_{3,u}=e\displaystyle\frac{g_{NY_cD}}{u-M^2}|\vec 
q|\displaystyle\frac{W-m}{W+m}
\left [ 2Q_{Y_c}+\kappa_c\displaystyle\frac{W+m}{M}\right ],$$
$$f_{4,u}=e\displaystyle\frac{g_{NY_cD}}{u-M^2}(E_2-M)\left [- 
2Q_{Y_c}+\kappa_c\displaystyle\frac{W-m}{M}\right ],$$
where $W^2=s,$ $E_2=(s+M^2-m_D^2)/(2W)$
\item $t$-channel (D-contribution):

$$f_{1,t}(D)=f_{2,t}(D)=0,$$
$$f_{3,t}(D)=-2eQ_D\displaystyle\frac{g_{NY_cD}}{t-m^2_D}|\vec 
q|\displaystyle\frac{W-m}{W+m},$$
$$f_{4,u}(D)=2eQ_D\displaystyle\frac{g_{NY_cD}}{t-m^2_D}(E_2-M),$$

\item $t$-channel ($D^*$-contribution):
$$f_{1,t}(D^*)=\displaystyle\frac{e}{2}{\cal N}(g_1+g_2)
\left [t-m^2_D+2(W-m)(W-M)\right ]$$
$$+\displaystyle\frac{g_2}{m+M}
\left [-tW+mm^2_D-2(m+M)(W-m)(W-M) \right],$$
$$
f_{2,t}(D^*)=e{\cal N}(g_1+g_2)
\displaystyle\frac{W-m}{W+m}\displaystyle\frac{|\vec q|}{E_2+M}
\left\{ \displaystyle\frac{1}{2}
\left [t-m^2_D+2(W+m)(W+M)\right ]\right .$$
$$+\left .\displaystyle\frac{g_2}{m+M}
\left [tW+mm^2_D-2(m+M)(W+m)(W+M)\right] \right \},$$

$$f_{3,t}(D^*)=-e{\cal N}|\vec q|(W-m)\left [g_1+g_2-g_2\displaystyle\frac{W+m}{W-m}\right ],$$
$$f_{4,t}(D^*)=-e{\cal N}(W-m)(E_2+M)\left [g_1+g_2+g_2\displaystyle\frac{W-m}{W+m}\right ],$$
\end{itemize}
with $${\cal N}=\displaystyle\frac{g_{D^*D\gamma}}{(t-m^2_{D^*})}
\displaystyle\frac{1}{m_{D^*}}.$$

{}

\end{document}